\documentclass{ametsocV6.1}

\usepackage{physics}
\usepackage{cancel}
\newcommand{\rhobar}{\overline{\rho}}

\newcommand{\irhobar}{\frac{1}{\overline{\rho}}}

\title{Multi-scale dynamics of deep convective self-aggregation. Part I: physical-space structures and energy variance budget}

\authors{Lorenzo Silvestri,\aff{a}\correspondingauthor{Lorenzo Silvestri, lorenzo.silvestri@unipg.it} 
Paolina Bongioannini Cerlini,\aff{b} 
Andrea Cimarelli\aff{c} 
}

\affiliation{\aff{a}{Department of Civil and Environmental Engineering, University of Perugia}\\
\aff{b}{Department of Physics and Geology, University of Perugia}\\
\aff{c}{Department of Engineering "Enzo Ferrari", University of Modena and Reggio Emilia}
}

\abstract{Vertically resolved budget analyses of frozen moist static energy (FMSE) variance within a cloud-resolving model provide a rigorous perspective on the equilibrium state of deep convective self-aggregation. The analysis reveals that the variance budget is governed by a leading-order balance between vertical variance advection and the gradient term, which quantifies the coupling between vertical covariance FMSE fluxes and the mean vertical stratification. Because vertical FMSE covariance fluxes are strictly positive throughout the atmospheric column, the sign of the gradient term is entirely dictated by the vertical gradient of mean FMSE. The primary peak of variance production occurs at the boundary layer top, driven by the interplay between sharp FMSE gradients — intensified in dry regions where the moist boundary layer decouples from the dry free troposphere — and robust vertical covariance fluxes. The persistent positivity of these fluxes implies that moist (dry), high-energy (low-energy) anomalies are systematically tied to updrafts (downdrafts), regardless of the mean gradient sign. This breakdown of classical down-gradient mixing indicates that vertical transports are governed by non-local, long-distance interactions driven by large-scale overturning circulation. The large variance produced near the boundary layer top is subsequently transported upward and dissipated in the upper troposphere, where the gradient term turns into a sink. While radiation does not directly force local variance, it indirectly shapes the budget by establishing the mean stratification. Overall, local thermodynamic forcings are translated by the overturning circulation into the spatial redistribution of variance across multiple scales. This multiscale dynamical organization provides the physical foundation for Part 2 of this work, which adopts a scale-by-scale formalism to examine how locally generated variance cascades through the system.\\} 

\begin{document}

\maketitle

%
%
%
\statement
Understanding how tropical convection organizes itself into large, persistent cloud clusters is crucial for improving climate models and weather predictions. This study reveals the physical engine behind this process by tracking the spatial variance of atmospheric energy. We show that energy variance is primarily generated at the top of the atmospheric boundary layer, driven by large-scale circulation that connect moist and dry regions rather than local turbulent mixing. By demonstrating that traditional mixing assumptions fail to capture these long-distance transport mechanisms, our findings provide a new physical framework for how convective systems redistribute energy across spatial scales. This paves the way for better parameterizations of convective aggregation in global climate models.
%
%

%

\section{Introduction}

The multiscale nature of cloud aggregation is a common feature of our planet. One representative example can be found in the tropics where convection and clouds tend to organize in a narrow band called Inter-Tropical Convergence Zone (ITCZ). From the point of view of large planetary scales, the ITCZ looks as the humid and ascending branch of the Hadley cell. Its width may vary with latitude, may assume a single or dual-band form, may be weaker or stronger subjected to aggregation and disaggregation cycles \citep{Maithel2024aggcycles} or may even break down \citep{Casallas2025breakdown}. Meridional gradients of moisture build up during aggregation cycles mainly as a consequence of horizontal advection of moist static energy and an anomalous flow of moist air from dry to moist columns at low levels \citep{Maithel2024aggcycles}. A recent theory on moisture modes from \cite{Adames2024stirring} suggests that these meridional gradients of moisture may become unstable and act as source of waves which stirs humidity from the tropics toward the mid-latitudes, leaving behind a more homogeneous humid state. A recent work by \cite{tompkins2025drivers} found verification of this theory in the tropical western pacific, by highlighting the effect of westward propagating waves in disrupting aggregated pattern of convection. This confirm the fundamental role in horizontal advection of moisture in determining when the domain aggregates and disaggregates \citep{Maithel2024aggcycles} and that large-scale processes have an important role in the disruption of large-scale aggregation. However, its development during aggregation cycles has to be reconducted to a strong interaction of convection across scales, where also local gradients of moisture at small scale can contribute to the creation of large moisture modes at planetary scales.

Idealized simulations of deep convective self-aggregation under radiative-convective equilibrium and without external forcings are a perfect setup to study the role of scale interaction during aggregating cycles. In a domain large enough (at least one dimension bigger than 100 km), with lateral periodic boundary condition and uniform Sea Surface Temperature (SST), initial random convection can self-aggregate in a single cluster with a time scale of order 10 days. Self-aggregation can occur in similar setup also in shallow cumulus simulations \citep{Janssens2024shallow,Bretherton2017mesoscaleagg}. Its occurrence in cloud permitting models with horizontal resolutions of the order of 1 km has been extensively studied in the literature starting from the seminal works of \cite{tompkins1998radiative,tompkins2001watervapor,tompkins2001coldpools, bretherton2005energy, muller2012detailed, wing2014physical} up to more recent model-intercomparison in the RCEMIP project \citep{wing2020clouds}, where the spontaneous aggregation of clouds has been recognized to a be a robust feature across different cloud permitting models. The role of the different processes (radiative, surface fluxes, advective) in numerical simulation of cloud aggregation has been summarized in recent reviews by \cite{wing2018convective} and \cite{muller2022spontaneous}. Different experiments have shown the sensitivity of self-aggregation to large-scale mechanisms such as radiative anomalies \citep{Muller2015whatfavors}, mean wind \citep{jung2021convective}, wind shear \citep{tompkins2001watervapor,bretherton2005energy}, rotation \citep{carstens2022spectrum}, different Sea Surface Temperature \citep{shamekh2020self,wing2020clouds}, sub-cloud layer properties \citep{cerlini2023competing} and different domain sizes \citep{muller2012detailed,Yanase2020critical, patrizio2019sensitivity}. Many others have shown its sensitivity to small-scale phenomena and physical parametrization such as turbulent mixing \citep{tompkins2017organization,silvestri2024numerical}, planetary boundary layer and microphysics schemes \citep{shi2021modulation,casallas2025sensitivity}. The strong interaction of such mechanisms at different scales is also highlighted by the relevance to this problem of different structures at different scales: low clouds and shallow circulations \citep{muller2012detailed,shamekh2020self}, cold pools and convective updrafts \citep{jeevanjee2013convective,tompkins2017organization}, dry plumes \citep{yang2020initiation},  large-scale cluster density anomalies \citep{windmiller2019convection} and large scale moist static energy circulation \citep{bretherton2005energy}.

The occurrence of self-aggregation is associated to an increase of the spatial variance of Frozen Moist Static Energy (FMSE). Hence, most studies adpoted the framework of a vertically integrated FMSE variance budget \citep{wing2014physical} in order to understand which components of the budget (radiative, advective or surface  fluxes) support or contrast self-aggregation during its temporal development. In radiative convective equilibrium simulations at the tropics, where gravity waves by deep convection efficiently remove any horizontal temperature gradient in the free troposphere \citep{bretherton1989gravity,sobel2001weak}, the increase in FMSE variance is mostly associated with an increase in water vapor variance. Therefore, the vertical integrated framework helps identifying the overall effect of each process on whole atmospheric column water vapor content, giving a global view of which processes tend to sustain (suppress) aggregation by moistening (drying) the moist columns and drying (moistening) the dry columns. However, within this approach is difficult to disentangle the role played by the different convective structures found at different scales and elevation above ground. \cite{yao2022vertically} introduced a vertically resolved framework by first evaluating the variance at each level and then integrating between different heights. This framework helped to verify previous hypothesis by \cite{Muller2015whatfavors,yang2018boundary} on the key importance of boundary layer and lower-free troposphere on self-aggregation. Indeed works by \cite{yang2018boundary,shamekh2020self} showed that the differential cloud radiative longwave cooling, often found to be a key process for self-aggregation development, actually acts through its influence on the spatial gradients of buoyancy in the planetary boundary layer. Therefore the vertical structure in self-aggregation is relevant as also demonstrated by the sensitivity of numerical simulations to vertical resolution \citep{jenney2023vertical}.

The study of vertically resolved energy variance budget is certainly not new either to the broader fluid dynamic community or to the atmospheric studies regarding boundary layer clouds \citep{deardorff1974three,de2004large,schemann2017budget}. \cite{khairoutdinov2002similarity} were the first to systematically study the budgets of second-order statistical moments such as kinetic energy, liquid/ice water static energy ($h_L$) and total water content ($q_T$) variances in continental deep cumulus convection. Their analysis highlights the dominance of gradient production terms, $2\overline{w'Y'}d\overline{Y}/dz$ (correlation between large-scale mean vertical gradient of scalar $Y$ and fluctuating vertical fluxes of $Y$), which in their study were mostly balanced by the precipitation related terms. Such production by large scale gradients remain hidden in the approach by \cite{yao2022vertically} which studied the adiabatic term as a whole, without separating it into a convective transport and a gradient production.

In this work, we will show that the study of the FMSE energy variance budget in deep convective self-aggregation, as done by using the classical framework of boundary layer studies, can reveal and explain important features about the multi-scale properties of convection. These results represent the first part of a research activity that includes also a second part \citep{silvestri2026gye_part2} where we will demonstrate how useful can be such a framework when transformed into a scale-space formalism, highlighting the role of each process at each scale. The paper is organized as follows: Section \ref{sec:eq} describes the governing equations of the numerical model and the flow setup of the RCE simulation; Section \ref{sec:rce} shows the obtained radiative-convective equilibrium and the steady state of convection; Section \ref{sec:vertical_profiles} shows the vertical inhomogeneity of the equilibrium states; Section \ref{sec:condprof} analyzes the conditionally averaged statistics to show the strongly organized structures of self-aggregation in the horizontal direction; Section \ref{sec:variance} shows the energy variance budget and which terms are most relevant for the variance production during the maintenance stage of self-aggregation; Section \ref{sec:conclusions} provides the discussion about the results and concludes this work.

\section{Governing equations and flow problem}
\label{sec:eq}

The System of Atmospheric Modeling \citep[SAM version 6.10.5,][]{khairoutdinov2003cloud} is employed here to study convective self-aggregation. This model adopt the so-called anelastic approximation, assuming a fixed reference profile of density which varies with height $\overline\rho=\overline\rho(z)$ and fulfill the hydrostatic equilibrium, where the operator $\overline{(\cdot)}$ denotes the average along the lateral (horizontal) directions. The mass conservation equation then reads:
\begin{equation}
    \pdv{\overline{\rho} u_i}{x_i}=0 \qquad  \pdv{u_i}{x_i}=-\frac{w}{\overline{\rho}}\frac{d\overline{\rho}}{dz}
    \label{eq:mass}
\end{equation}
The momentum conservation is:
\begin{equation}
    \frac{\partial u_i}{\partial t} + \frac{1}{\overline{\rho}}\frac{\partial \overline{\rho}u_i u_j}{\partial x_j}=-\frac{\partial}{\partial x_i} \left (\frac{p''}{\overline{\rho}} \right )-\delta_{i3} g \frac{\rho''}{\overline{\rho}} \,- \frac{1}{\overline{\rho}}\frac{\partial \tau_{ij}}{\partial x_j}+\left(\frac{\partial u_i}{\partial t} \right)_{LS}
    \label{eq:momentum}
\end{equation}
where zero Coriolis force is assumed as in the tropical atmosphere. The terms $(p^{\prime\prime},\rho^{\prime\prime})$ represent the fluctuations with respect to the horizontal mean $(\overline{p},\overline{\rho})$, respectively. The buoyancy term is approximated as:
\begin{equation}
    B=-g \frac{\rho''}{\overline{\rho}} \approx g \left (\frac{T''}{\overline{T}}+0.608 q_v''-q''_{liq}-q''_{ice} \right )
\end{equation}
where $T, q_v, q_{liq}, q_{ice}$ denotes absolute temperature, water vapor mixing ratio, liquid water mixing ratio and ice water mixing ratio. The reference temperature profile over which temperature perturbation is evaluated (as well as the perturbations on $q_v,q_{liq},q_{ice}$), is updated each time step, while density profile is kept constant. Therefore this buoyancy formulation assumes negligible contribution of pressure perturbations with respect to temperature ones \citep{khairoutdinov2022global}. 
Liquid water and ice water are defined as: 
\begin{align}
    q_{liq} &= q_r+q_{cl} \\
    q_{ice} &= q_s+q_g+q_{ci}
\end{align}
where $q_r, q_s, q_g, q_{cl}, q_{ci}$ are respectively rain water mixing ratio, snow water mixing ratio, graupel water mixing ratio, cloud water mixing ratio and cloud ice mixing ratio.
The last term on the RHS of Eq. \ref{eq:momentum} is an additional large-scale momentum forcing. In our simulation, as in previous studies \citep{muller2012detailed}, there is no imposed mean wind, but the horizontally averaged winds are relaxed over a time scale of 1h toward zero at all vertical levels. Finally, $\tau_{ij}$ are the modelled subgrid stresses.

Following the work by \cite{pauluis2008anelastic} for the energy conservation under the anelastic approximation, the thermodynamic energy conserved in SAM is the Frozen Moist Static Energy (FMSE) evaluated at the reference state density, $\overline{\rho}(z)$, and defined as:
\begin{equation}
    h_f=c_{pd}T  + gz + L_c q_v - L_f q_{ice}.
    \label{eq:def_hf}
\end{equation}
where $c_{pd},L_c,L_f$ are the specific heat of dry air, the latent heat of condensation and latent heat of fusion. $h_f$ represents the sum of internal energy, potential energy and the latent energy associated to water vapor and water ice. In Equation \ref{eq:def_hf}, the latent component of energy is defined as to represent phase changes relative to liquid water. Water vapor condensation into liquid water ($L_c q_v$) contributes positively to the energy since it releases sensible heat, while water ice fusion into liquid water ($-L_f q_{ice}$) contributes negatively to the energy since it absorbs sensible heat. Conservation of $h_f$ requires prognostic equations for $T$ and water species $q_v,q_{ice},q_{liq}$. In SAM the thermodynamic variable used is the liquid-ice water static energy, $h_L$, which is derived from $h_f$ by separating off the latent energy component:
\begin{equation}
    h_L = h_f-L_c(q_v+q_{liq}+q_{ice})=c_{pd}T  + gz - L_c q_{liq} - L_s q_{ice}
\end{equation}
which represents the sum of internal energy, potential energy and the latent energy of liquid and ice water particles (written with respect to water vapor phase). $L_s=L_c+L_f$ is the latent heat of sublimation. The conservation equation of $h_L$ is:
\begin{equation}
    \frac{\partial h_L}{\partial t} + \frac{1}{\overline{\rho}}\frac{\partial \overline{\rho}u_j h_L}{\partial x_j} =-\frac{1}{\overline{\rho}}\frac{\partial}{\partial z}(L_c P_{liq}+L_s P_{ice})- \frac{1}{\overline{\rho}}\frac{\partial \tau_{h_L, j}}{\partial x_j} + c_{pd}Q_{rad}\\
    \label{eq:hl}
\end{equation}
where $P_{liq}=P_r$ is the liquid precipitation fluxes (rain), while $P_{ice}=P_s+P_g+P_i$ are the frozen precipitation fluxes (snow, graupel and ice sedimentation), $\tau_{h_L j}$ are the sub-grid scalar fluxes and $Q_{rad}$ is the radiative heating rate in $K/s$. Lastly we need other two scalar conservation equations for the total non-precipitating water $q_{np}= q_v+q_{cl}+q_{ci}$ (water vapor, cloud water and cloud ice) and the total precipitating water $q_p = q_r+q_s+q_g$ (rain, snow, graupel):
\begin{align}
    &\frac{\partial q_{np}}{\partial t} + \frac{1}{\overline{\rho}}\frac{\partial \overline{\rho}u_j q_{np}}{\partial x_j} =\frac{1}{\overline{\rho}}\frac{\partial P_i}{\partial z}-\frac{1}{\overline{\rho}}\frac{\partial \tau_{q_{np} j}}{\partial x_j} - \left(\frac{\partial q_{p}}{\partial t} \right)_{mic}
    \label{eq:qnp} \\
     &\frac{\partial q_p}{\partial t} + \frac{1}{\overline{\rho}}\frac{\partial \overline{\rho}u_j q_p}{\partial x_j} =\frac{1}{\overline{\rho}}\frac{\partial}{\partial z}(P_r+ P_s + P_g)- \frac{1}{\overline{\rho}}\frac{\partial \tau_{q_{p} j}}{\partial x_j} + \left(\frac{\partial q_{p}}{\partial t} \right)_{mic} \label{eq:qp}
\end{align}
where $\tau_{q_{p} j}$ and $\tau_{q_{np} j}$ are the sub-grid scalar fluxes and the last term on the RHS of both equations is the source/sink due to the microphysics which controls the phase changes between the different water species and the precipitation fluxes. It is important to recall here that cloud ice, $q_{ci}$, although defined as a non-precipitating water substance, is allowed to fall at a terminal velocity which is dependent on the cloud ice content itself \citep{Heymsfield2003vterminal}. This process, which we referred to as ice sedimentation, also affect $h_L$ (Eq.\ref{eq:hl}) by taking into account the latent heat flux induced by the falling cloud ice ($P_i$). We can now write the evolution equation for the total water $q_T = q_{np}+q_p = q_v + q_{liq}+q_{ice}$ by combining equations \ref{eq:qnp} and \ref{eq:qp}:
\begin{equation}
     \frac{\partial q_T}{\partial t} + \frac{1}{\overline{\rho}}\frac{\partial \overline{\rho}u_j q_T}{\partial x_j} = \frac{1}{\overline{\rho}}\frac{\partial}{\partial z}(P_{liq}+P_{ice})- \frac{1}{\overline{\rho}}\frac{\partial \tau_{q_{T} j}}{\partial x_j}
     \label{eq:qt}
\end{equation}
Since $h_f=(h_L+L_c q_T)$, we multiply Eq. \ref{eq:qt} by $L_c$ and add it to Eq. \ref{eq:hl} to find an equation for the frozen moist static energy per unit volume $\overline{\rho} h_f$:
\begin{equation}
    \frac{\partial \overline{\rho} h_f}{\partial t} + \frac{\partial \overline{\rho}u_j h_f}{\partial x_j} =-\frac{\partial L_f P_{ice}}{\partial z}- \frac{\partial \tau_{h_f j}}{\partial x_j} + \overline{\rho} c_{pd}Q_{rad}\\
    \label{eq:hf}
\end{equation}
where we have renamed the frozen precipitation fluxes $P_{ice}=P_f$ (to have a direct link to the notation of $h_f$) and $\tau_{h_f j}=\tau_{h_L j}+L_c \tau_{q_T j}$.
Sub-grid momentum stresses and sub-grid scale scalar fluxes are modeled by using the Smagorinsky model. Radiation is interactive and CAM3 parametrization is used for radiative heating, considering that the domain is centered at zero latitude and the diurnal cycle is present. For microphysics, the original SAM model (SAM1MOM) is used.

The numerical simulation is performed over a squared domain of dimensions $(L_x,L_y,L_z)=(768,768,27)km$, where $x$ and $y$ are the lateral directions and $z$ is the vertical direction (aspect ratio $A=L_z/L_x\approx$ 28). Lateral boundary conditions are periodic. At the bottom, surface momentum and scalar fluxes (e.g. temperature, moisture) are prescribed by using Monin-Obukhov Similarity Theory and assuming a water surface with a constant sea surface temperature of 302 K and surface pressure of 1015 hPa. The first model level is at 25 m, while a rigid lid is present at the top at about 27 km. To prevent unphysical wave reflection on the upper boundary, we use a sponge layer with traditional Rayleigh damping starting at the upper third of model domain (18 km).

The SAM anelastic equations are integrated in time with an explicit 3rd order Adam-Bashfort scheme. Momentum advection is performed by using a 2nd order central difference scheme in a flux form for kinetic energy conservation. Scalar advection is based on the 5th order monotonic ULTIMATE-MACHO scheme as developed by \citet{yamaguchi2011cloud}. The model uses a staggered Arakawa C-type grid with uniform spacing of elements in the lateral direction and a stretched spacing in the vertical one. The total number of grid elements is $(N_x,N_y,N_z)=(256,256,64)$. The resulting horizontal grid resolution is $\Delta x=\Delta y=3$ km, small enough to consider the simulation "convective and cloud permitting". In the vertical direction, the grid spacing gradually increases from 50 m near the ground to 500 m above 5 km. Then it increases again from 500 m to 1 km above 20 km.

The simulation is initialized from two equilibrium profiles of temperature and water vapor of a corresponding smaller domain RCE simulation (averaged over the last 20 days) and convection is triggered by adding white noise to $h_L$ in the lowest five model levels, with an amplitude of 0.1 K in the lowest level linearly decreasing to 0.02 K in the fifth level.

The symmetries of our problem (lateral periodic boundary conditions) are such that the solution is statistically inhomogeneous only in the vertical direction $z$. The average operator $\langle\, \cdot \,\rangle$ is used to denote an average both in time (last 10 days) and space (only horizontal directions). Instead, the average operator $\overline{\;\cdot\;}$ is used to denote only an average along the horizontal directions. The fluctuation of a certain quantity $Y(x,y,z,t)$ with respect to both space and time average is indicated as $Y^{\prime}(x,y,z,t)=Y(x,y,z,t)-\langle Y \rangle (z)$, while the fluctuation with respect only to space is indicated as $Y^{\prime\prime}(x,y,z,t)=Y(x,y,z,t)-\overline{Y}(z,t)$. Averages in time will be performed over the last 10 days, for a total of 240 samples.

\section{Temporal evolution and the statistical radiative-convective equilibrium}
\label{sec:rce} 

\begin{figure}[t]
    \centering
    \includegraphics[width=1\linewidth]{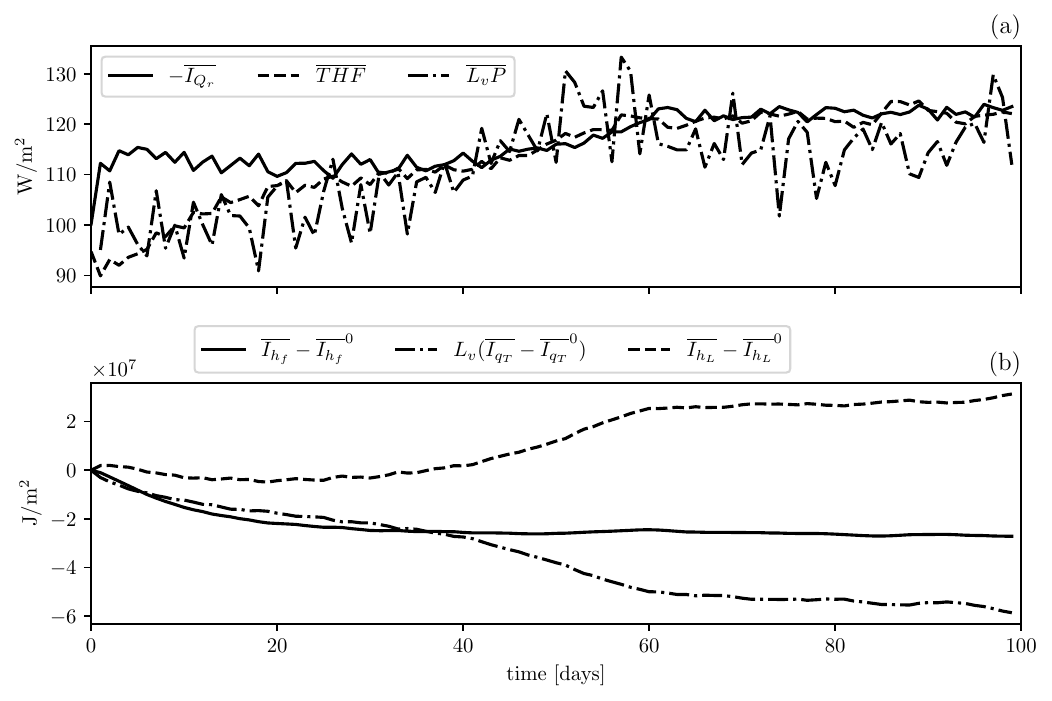}
    \caption{a) Time evolution of domain averaged of the the column integral of radiative heating rate, $I_{Q_r}$, and surface total heat fluxes, $THF$, as well as the latent energy associated to surface total precipitation fluxes, $L_v P$. b) Time evolution of the anomaly of the column integral of frozen moist static energy $I_{h_f}$ with respect to its initial value togheter with its two component ($I_{q_T}$ and $I_{h_L}$). Values are daily averaged and domain averaged.}
    \label{fig:fmse_int_evo}
\end{figure}

The statistical atmospheric condition known as the radiative-convective equilibrium can be identified by studying the evolution of Eq. \ref{eq:hf} once integrated over the vertical direction:
\begin{equation}
    \frac{\partial I_{h_f}}{\partial t} = -L_f [P_{f}]_{z_0} - \int_{z_0}^{z_t}\overline{\rho} \frac{\partial \tau_{h_f,j}}{\partial x_j}dz + \int_{z_0}^{z_t}\overline{\rho}c_{pd}Q_{rad}dz
\end{equation}
where $z_0$ and $z_t$ are respectively the height of bottom model level (25 m) and top model level (27 km) above ground and 
\begin{equation}
    I_{h_f}=\int_{z_0}^{z_t}\overline{\rho} h_f dz
\end{equation}
is the the column integral of frozen moist static energy. By considering the lateral periodic boundary conditions (no incoming or outgoing fluxes of heat or moisture), considering zero fluxes of sensible heat and moisture at the top of the domain, and averaging along the horizontal directions we obtain:
\begin{equation}
    \frac{d \overline{I_{h_f}}}{dt}  \approx - \underbrace{\overline{\rho}[\overline{\tau_{h_f,z}}]_{z_0}}_{\overline{THF}} + \underbrace{\int_{z_0}^{z_t}\overline{\rho}c_{pd}\overline{Q_{rad}}dz}_{\overline{I_{Q_r}}}
\end{equation}
where the contribution of latent energy lost by surface frozen precipitation fluxes (graupel and snow) has been also neglected. Under steady equilibrium we recover the balance between total averaged surface heat fluxes ($\overline{THF}$) and total averaged net column radiative heating ($\overline{I_{Q_r}}$). Such integral quantities are shown on Figure \ref{fig:fmse_int_evo}a. From Figure \ref{fig:fmse_int_evo}a it is evident that, even if starting from a closely balanced situation, the flow reaches a steady equilibrium only after 60 days.

The evolution of the column integral of frozen moist static energy can be decomposed in its two contributions $\overline{I_{h_f}}= \overline{I_{h_L}} + \overline{I_{q_T}}$, where $I_{h_L}=\int_{z_0}^{z_t}\overline{\rho} h_L dz$ is the column integral of liquid ice static energy, and $I_{q_T}=\int_{z_0}^{z_t}\overline{\rho} L_c q_T dz$ is the latent heating energy associated to the column integral of water content. As shown in Figure \ref{fig:fmse_int_evo}b, such decomposition highlights how the average energy decreases over the domain because the decrease in the total water content is larger than the increase in total dry static energy content. In other words, the domain dries out during all the simulation. Instead, the warming is mostly concentrated after 40 days. Before such warming, the domain has constant internal energy, or slightly decreasing. 

\begin{figure}[t]
    \centering
    \includegraphics[width=0.8\linewidth]{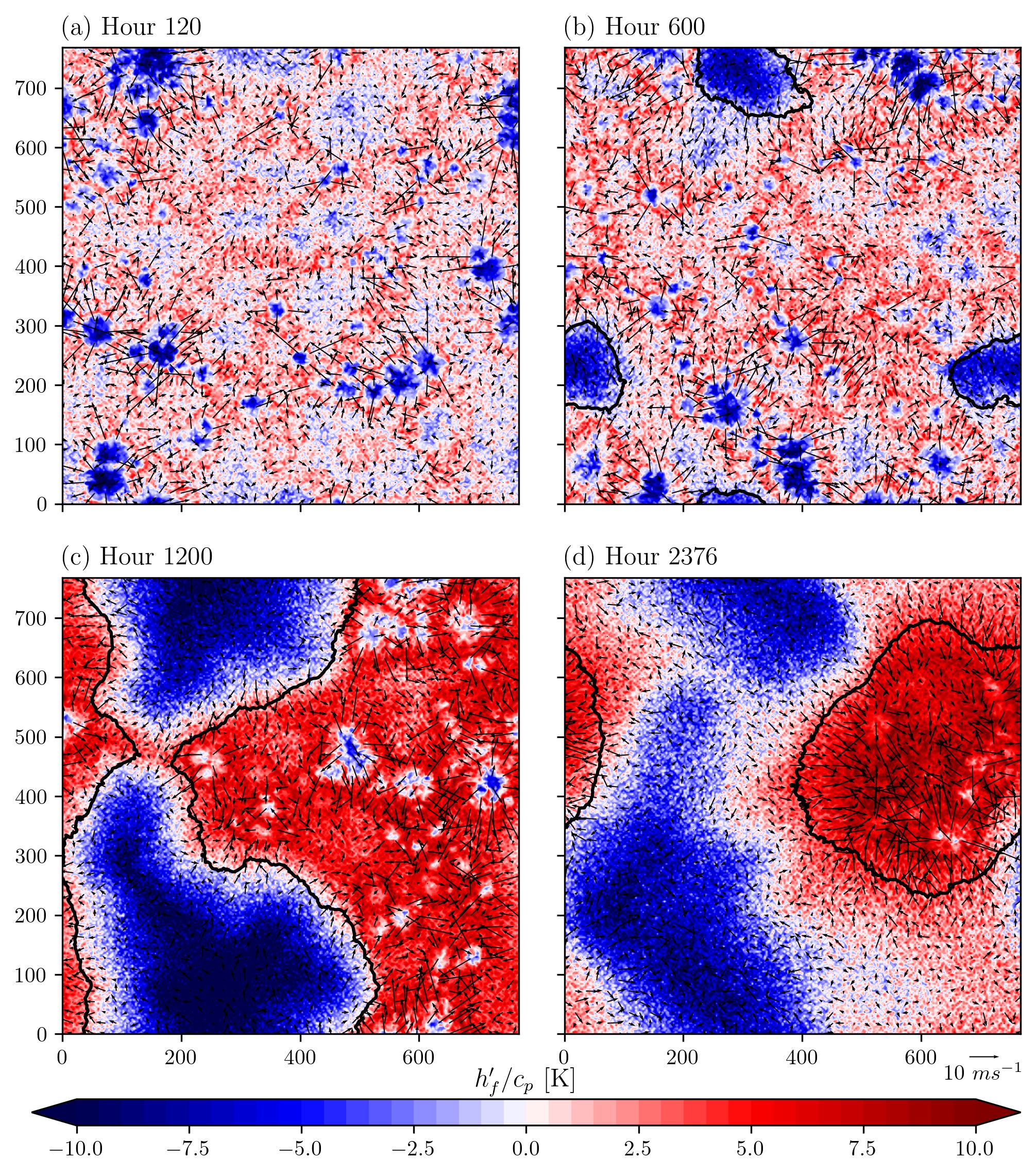}
    \caption{Snapshots of $h_f''=h_f-\overline{h_f}$ at the first model level (surface) after a) 5 days, b) 25 days, c) 50 days, d) 99 days. Black arrows indicates surface wind (see the magnitude reference on the bottom right). Solid black contour line indicates where Column Relative Humidity (CRH) is 0.5 (dry regions).}
    \label{fig:snap_hfprime}
\end{figure}

The analysis of the temporal evolution of the atmospheric system can now be expanded by considering also the evolution of the structure taken by the solution in the horizontal directions. To this aim, the evolution of the anomaly of frozen moist static energy is reported in Figure \ref{fig:snap_hfprime} where dry and moist convective regions are also identified by considering the Column Relative Humidity, defined as 
\begin{equation}
    CRH=\frac{\int_{z_0}^{z_t}\overline{\rho} q_v dz}{\int_{z_0}^{z_t}\overline{\rho} q_{v,sat} dz}
\end{equation}
that is the column integral of total water vapor divided by the column integral of total saturation water vapor that depends on the temperature and pressure of each cell. Such identification is based on the exponential relationship that has been observed between precipitation and CRH \citep{Bretherton2004,rushley:18}: $P(CRH)=P_0 e^{a_d CRH}$, with $P_0$ and $a_d$ being constant values. The value of $\langle CRH \rangle$ over the last 10 days is about 0.41. Therefore, the boundary between moistest and dry region can be identified as CRH $=$0.5 and is reported with a solid black contour in Figure \ref{fig:snap_hfprime}.

As shown in Figure \ref{fig:snap_hfprime}a, the initial stages of the simulation are characterized by random convection throughout the domain, replicating what is found in the small domain simulations used as initial conditions. 
During the first 5/10 days, deep convection is randomly distributed across the domain, as well as the associated cold pool (negative values of surface $h_f$ anomaly) and downdrafts, which triggers new convection over all the domain. After 20 days some very dry regions have formed throughout the atmospheric columns. 
The dry regions that exhibit $CRH < 0.5$ persist in time (not shown) and are free of convection. Multiple dry patches forms during this period and expand until after 40 days they start to merge together and bound convection into a single cluster which remains steady (Figure \ref{fig:snap_hfprime}d). Radiative convective equilibrium, $\overline{THF}\approx-\overline{I_{Q_r}}$ is recovered only after dry patches start to merge together after 40 days as seen in Figure \ref{fig:fmse_int_evo}a. The same is true for the surface precipitation fluxes equilibrium $\overline{THF}\approx\overline{L_v P}$ (precipitation fluxes are defined as positive downward by convention). In the first phase, until 40 days, the net radiative cooling is larger than surface heat fluxes, while the moisture budget (between precipitation fluxes and surface latent heat fluxes) is mostly fulfilled. This energy deficit (larger cooling than heating), is used mostly to dry out the domain, while only a small part is use to decrease his integral liquid-ice static energy. The process of transition from random deep convection to a single cluster is what we called self-aggregation, since it is due to internal feedbacks.
Therefore, we can divide the time evolution of energy into four periods: creation of large dry patches (0-10 days), expansion of dry patches (10-40 days), merging (40 days-60 days) and steady state deep convective cluster. The final average daily precipitation rate, $\langle P \rangle$ is about 4 mm/day (or 118 W/$m^2$ in Figure \ref{fig:fmse_int_evo}a), which is also in line with the prediction by using the exponential relationship provided by \cite{Bretherton2004} ($P=P_0 e^{a_d\,CRH}$, with $P_0=4.07 \cdot 10^{-5}$ and $a_d=16.12$) and considering an average CRH$\approx 0.7$ on the moist regions.

\section{Vertical inhomogeneity: atmospheric layers}
\label{sec:vertical_profiles} 

As already shown in Section \ref{sec:eq}, the atmospheric settings are such that the solution is statistically non-homogeneous only in the vertical direction. In order to characterize the atmospheric layers that are developed in the final equilibrium stage of the simulation, we analyse here the vertical profiles and budgets of frozen moist static energy.

\begin{figure}[t]
    \centering
    \includegraphics[width=1\linewidth]{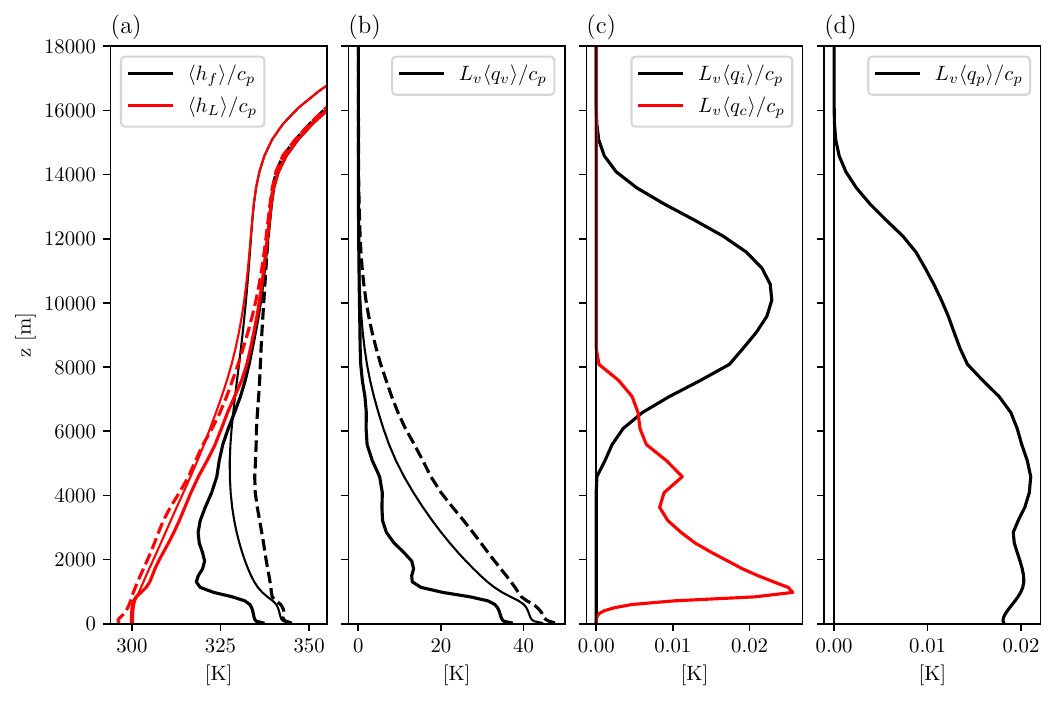}
    \caption{Vertical mean profiles over the last 10 days: a) $\langle h_f \rangle/c_p$ and $\langle h_L \rangle/c_p$, b)$L_v \langle q_v \rangle/c_p$ and c) $L_v \langle q_i \rangle/c_p$ and $L_v \langle q_c \rangle/c_p$, d)$L_v \langle q_p \rangle/c_p$ . Solid thick line indicates the average over the last 10 days, solid thin line is the initial condition, dashed lines are the profiles conditionally averaged inside clouds (region with positive $q_c+q_i$) for the last 10 days.}
    \label{fig:mean_prof}
\end{figure}

The vertical profiles of $\langle h_f \rangle$ are shown in Figure \ref{fig:mean_prof}a. Compared to the initial condition, $\langle h_f \rangle$ have decreased over the layers from the surface up to the mid-upper troposphere at about 6000 m, while the upper troposphere and lower stratosphere has warmed up. The liquid ice static energy has remained almost constant over the boundary layer up to 1000 m, while it increases above. The strong decrease of $\langle h_f \rangle$ is related to the expansion of large dry patches which reaches also the boundary layer, since under the cloud the boundary layer maintains almost the same $\langle h_f \rangle$ as during the initial time step. 

The conditional statistics of $\langle h_f \rangle$ over the clouds, defined as regions with $(q_c+q_i)>0$, showed that this variable is almost conserved on average from the boundary layer toward the upper troposphere, since inside clouds the increase of $h_L$ balances with the decreases of water vapor (Figure \ref{fig:mean_prof}b). The contribution of cloud ice, cloud water and precipitating water is very small (notice the different scale range in Figure \ref{fig:mean_prof}c and \ref{fig:mean_prof}d) and it reaches values of the order of 1 K only inside the cloud, with a peak at about 8000 m for cloud ice, about 3000 m for cloud water and about 4000 m for precipitating water (not shown). The vertical profiles of moisture highlights two region of strong gradient: the surface layer involving only the first 4 model levels up to 250 m, and the top of the boundary layer from 750 m to 1200 m. 

\begin{figure}[t]
    \centering
    \includegraphics[width=1\linewidth]{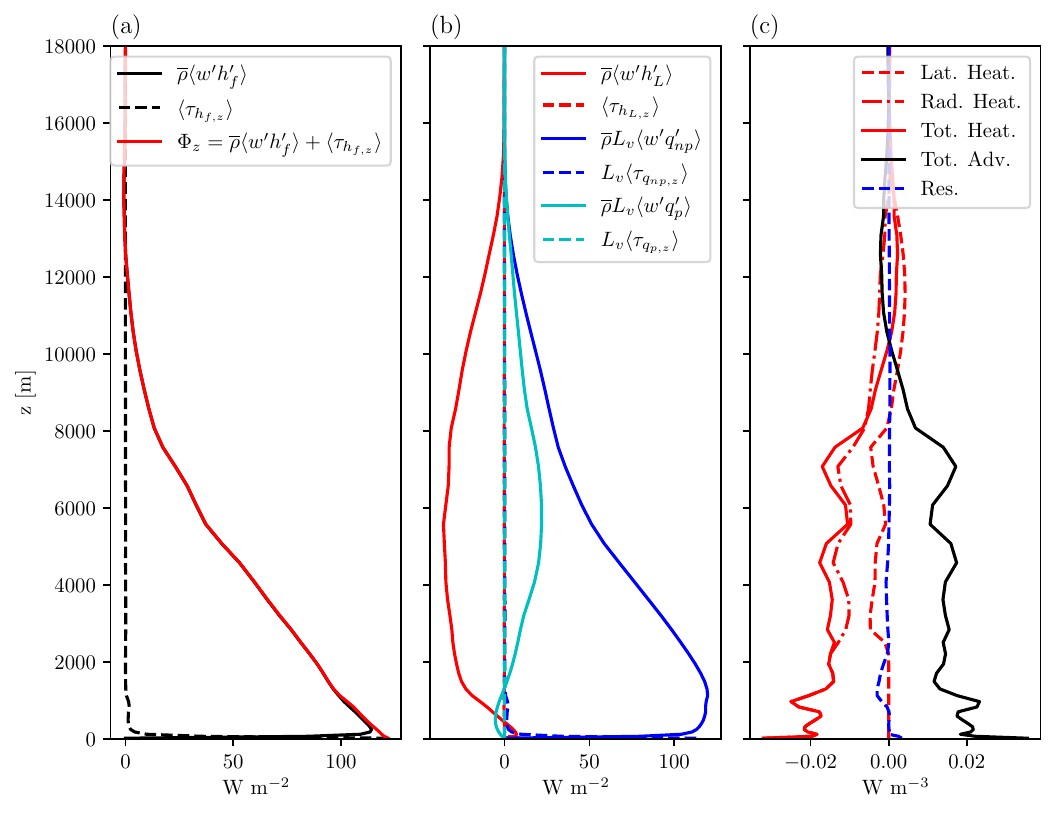}
    \caption{Mean fluxes and energy budget of $\langle h_f \rangle$: a) resolved and sub-grid scale fluxes of $h_f$, b) resolved and sub-grid scale fluxes components of $h_f$, c) energy budget of $h_f$ as reported in Eq \ref{eq:hf_budg_mean}, where Tot. Heat. refers to the total heating as the sum of terms Lat. Heat. and Rad. Heat. .}
    \label{fig:mean_flux}
\end{figure}

The budget of $\langle h_f \rangle$ can be obtained by averaging equation \ref{eq:hf} and reads:
\begin{equation}
    \underbrace{\frac{\partial \overline{\rho} \langle w^{\prime} h^{\prime}_f\rangle}{\partial z}}_{Adv.} = -\underbrace{L_f\frac{\partial \langle P_f \rangle}{\partial z}}_{Lat. Heat.}-\underbrace{\frac{\partial \langle \tau_{h_f,z}\rangle}{\partial z}}_{SGS}+\underbrace{\overline{\rho}c_{pd}\langle Q_{rad}\rangle}_{Rad. Heat.}\\
    \label{eq:hf_budg_mean0}
\end{equation}
where a statistical steady-state has been assumed and the identity $\langle w h_f \rangle = \langle w' h_f ' \rangle$ has been used to write the advective term. We can express Equation \ref{eq:hf_budg_mean0} also in a conservative form, 
\begin{equation}
    \frac{\partial \Phi_{h_f}}{\partial z} = S 
    \label{eq:hf_budg_mean}
\end{equation}
thus highlighting the presence of a vertical flux of frozen moist static energy,
\begin{equation}
    \Phi_{h_f}=\overline{\rho} \langle w' h_f' \rangle+\langle \tau_{h_f,z}\rangle \, ,
\end{equation}
that is driven by local source/sink mechanisms,
\begin{equation}
    S= -\underbrace{L_f\frac{\partial \langle P_f\rangle}{\partial z}}_{Lat. Heat.}+\underbrace{\overline{\rho}c_{pd}\langle Q_{rad}\rangle}_{Rad. Heat.} \, .
\end{equation}
As shown in Figure \ref{fig:mean_flux}a, the total flux $\Phi_{h_f}$ is always positive, thus transporting moist static energy from the surface up to the top layers of the atmosphere and is maximum at the surface and decreases to zero at about 12000 m. At the surface it coincides with the subgrid surface fluxes, $\Phi_{h_f} \approx \langle \tau_{h_f,z}\rangle$, while at higher elevations is exclusively performed by advection mechanisms, $\Phi_{h_f} \approx \overline{\rho} \langle w' h_f' \rangle$. By considering the frozen moist static energy decomposition $h_f = h_L + L_v (q_{np}+q_p)$, both subgrid and advective fluxes can be decomposed as $\langle \tau_{h_f,z} \rangle = \langle \tau_{h_L,z} \rangle + \langle \tau_{q_{np,z}} \rangle + \langle \tau_{q_{p,z}}\rangle$ and $\overline{\rho} \langle w' h_f' \rangle= \overline{\rho} \langle w' h_L' \rangle + L_v \overline{\rho} (\langle w' q_{np}' \rangle + \langle w' q_{p}'\rangle )$. As shown in Figure \ref{fig:mean_flux}b, this decomposition highlights that the subgrid fluxes dominating the surface layer are those of latent heat, being those of sensible heat one order of magnitude lower, i.e. $\langle \tau_{h_L,z}\rangle \ll L_v \langle \tau_{q_{np,z}}\rangle$. On the other hand, the main contribution to the advective fluxes dominating the atmosphere above the surface layer, is given by the non-precipitating water, $L_v \overline{\rho} \langle w' q_{np}'\rangle$ (mainly water vapor) which is always positive. Instead, the flux of liquid ice static energy, $\overline{\rho} \langle w' h_L'\rangle$, is positive only above the surface layer up to 250 m, meaning that on average, above the surface layer, a downward flux of liquid ice static energy occurs that also enters the boundary layer. Finally, the precipitating water flux, $L_v \overline{\rho} \langle w' q_{p}'\rangle$ (liquid and frozen), exhibits a double role. It contributes to the overall upward transport of frozen moist static energy above the boundary layer where condensation occurs reaching a maximum over the mid-troposphere at about 6000 m. On the contrary, it contrasts the overall upward flux in the boundary layer where it is negative. Since precipitating water is non-zero only under the main cloud cluster, this means that on average on the boundary layer under the cloud precipitation is associated with a downward flux of frozen moist static energy.

All terms contributing to the budget of $\langle h_f \rangle$, Eq.\ref{eq:hf_budg_mean}, are shown in Figure \ref{fig:mean_flux}c. We see the net cooling effect of radiation which cools the entire atmosphere up to 14000 m, and it balances with the positive upward transport of energy by convection. Latent cooling due to ice melting is also contributing between 2000 and 8000 m. Above 8000, where we also have the peak of cloud ice (see Figure \ref{fig:mean_prof}c), water freezing is heating the upper troposphere. This effect is overcoming the net cooling due to radiation and advection becomes negative to compensate such heating. Radiative-convective equilibrium is therefore fulfilled at each level of the atmosphere, where the net cooling by radiation is balanced by convection (transport of heat) and latent heating by frozen precipitation fluxes. In general, above the surface layer where subgrid fluxes plays a substantial role, we can write an approximate reduced energy budget for the lower troposphere up to 8000 m as:
\begin{equation}
    \frac{\partial \overline{\rho} \langle w' h_f' \rangle}{\partial z} \approx \overline{\rho}c_{pd}\langle Q_{rad}\rangle\\
    \label{eq:red_budg_lowt}
\end{equation}
and one for the upper troposphere above 8000 m by considering also the latent heating due to frozen precipitation fluxes:
\begin{equation}
    \frac{\partial}{\partial z}\Big(\overline{\rho} \langle w' h_f' \rangle + L_f\langle P_f\rangle \Big) = \overline{\rho}c_{pd}\langle Q_{rad}\rangle\\
    \label{eq:red_budg_uppt}
\end{equation}

Finally, from Figures \ref{fig:mean_prof} and \ref{fig:mean_flux} we can identify relavant atmospheric layers that show homogeneous properties, which we will use later to describe various atmospheric processes. 
Following the behavior of the vertical profile of $\langle h_f \rangle$, we can define the following regions:
\begin{enumerate}
    \item Near-surface layer: between 0 and 250 m, where subgrid fluxes play a significant role, there is a strong decrease of $\langle h_f \rangle$ and the mean vertical fluxes of $\langle h_L\rangle$ are positive;
    \item Boundary layer: between 250 m and 750 m, a well mixed region where the gradient of $\langle h_f \rangle$ is almost null and still no cloud condensate is present. The vertical fluxes of $\langle h_L \rangle$ turn negative in this region, while the fluxes of precipitating water are negative.
    \item Boundary layer top: between 750 m and 1200 m, where we find the strongest gradient of $\langle h_f \rangle$ and shallow clouds start to form and a strong radiative cooling is present.
    \item Shallow clouds layer: between 1200 m and 2000 m, where the non precipitating water flux becomes positive and the vertical profile $\langle h_f \rangle$ shows a small inversion (positive gradient).
    \item Lower troposphere above the shallow clouds: between 2000 m and 3000 m, where $\langle h_f \rangle$ decreases again with height and right before frozen precipitation fluxes start to increase and cloud ice starts to form. 
    \item Mid free troposphere from 3500 to 8000 m where the gradient of $\langle h_f \rangle$ is positive and almost constant and the frozen precipitation fluxes contributes negatively to the mean energy budget (cooling the atmosphere).
    \item Upper troposphere above 8000 m, where the frozen precipitation fluxes contribute substantially to the energy budget and turn to have a heating.
\end{enumerate}

\section{Horizontal inhomogeneity: dry and moist atmospheric structures}
\label{sec:condprof}

\begin{figure}[t!]
    \centering
    \includegraphics[width=0.9\linewidth]{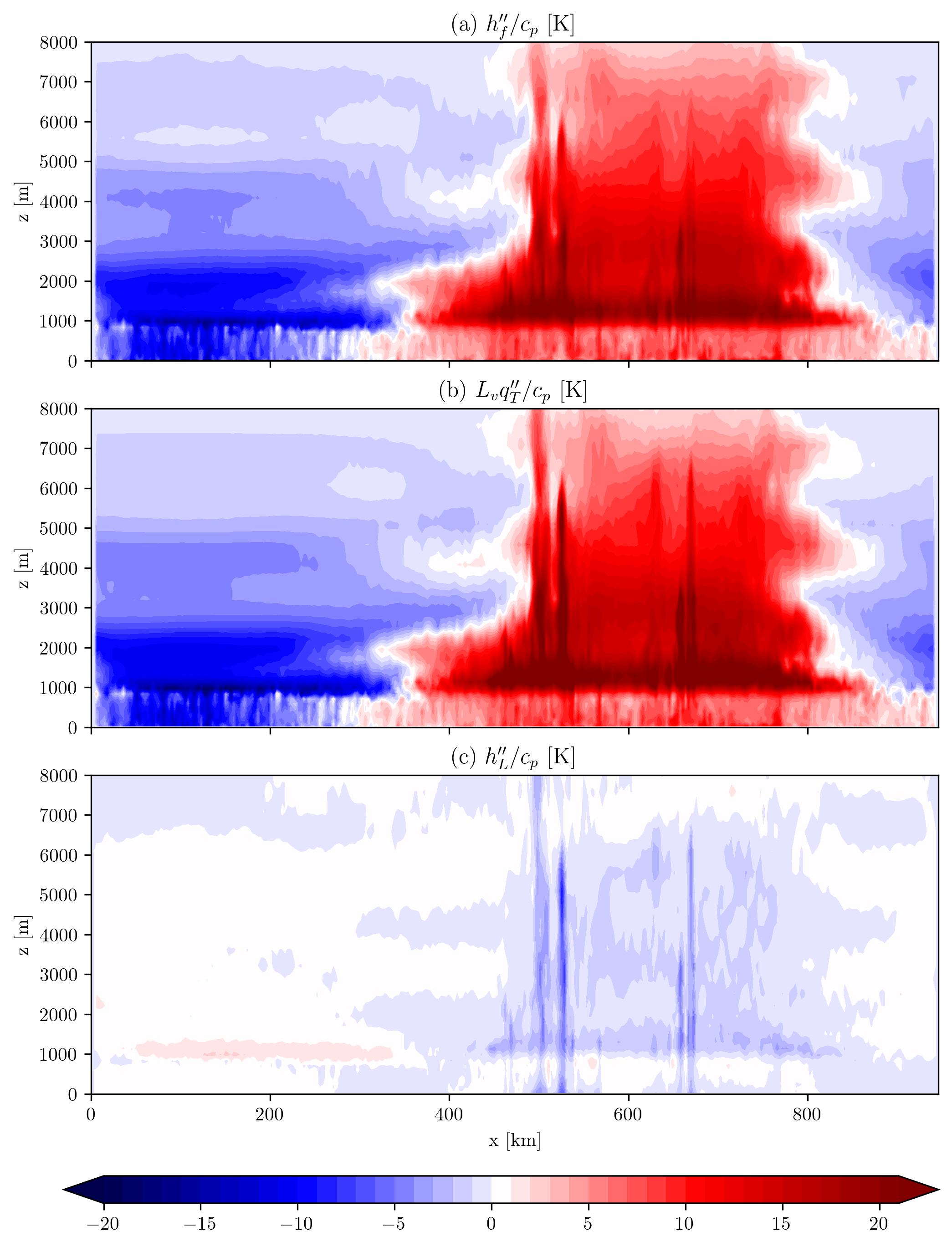}
    \caption{Snapshot of slice of $h_f''$ instantaneous fluctuations and its components over a diagonal line centered on the dry and moist region for the last hour of the simulation.}
    \label{fig:snap_slice}
\end{figure}

\begin{figure}[t!]
    \centering
    \includegraphics[width=1\linewidth]{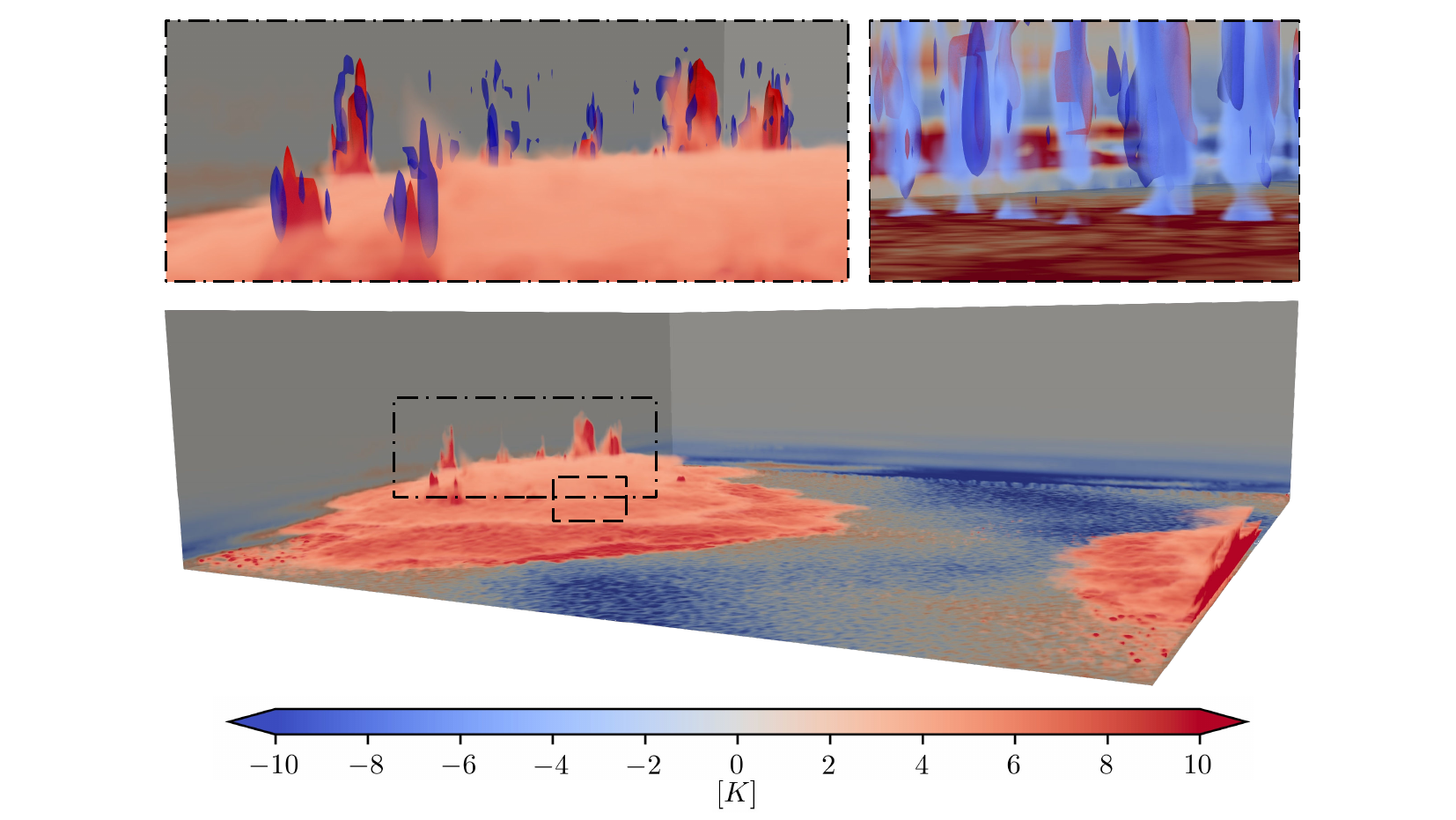}
    \put(-440,248){(a)}
    \put(-40,248){(b)}
    \put(-440,150){(c)}
    \caption{(c): Volume rendering of $L_v q_T''/c_p$ (transparent red indicates low values, while opaque red indicates high values) and slices of $L_v q_T''/c_p$ (blue indicates negative values and red indicates positive values). (a): Zoom on the top of the cloud to show deep convective towers. (b): Zoom inside the boundary layer below the deep convective activity to highlight cold pools shown as a volume rendering of $h_L''/c_p$ (from transparent light blue indicating high values to opaque blue indicating low values). In the zoomed figures (a) and (b), isosurfaces of positive and negative vertical velocity $w$ have been added ($w = 3 m/s$ in red and $w = -1.2 m/s$ in blue). The vertical dimension $z$ is elongated by 8 times with respect to the hoprizontal dimension (Aspect ratio vertical:horizontal=8:1).
    }
    \label{fig:3d}
\end{figure}

From the average budget and the statistics, we cannot appreciate the strong organization of the convective field and the non-homogeneity of the vertical structure of the atmosphere across the domain. In order to highlight the strong inhomogeneity of the solution, the main atmospheric structures are shown in Figure \ref{fig:snap_slice} through iso-contours of the frozen moist static energy anomaly, $h_f''=h_f-\overline{h_f}$. Note that for better visualization of the strongly anisotropic shape of the atmospheric structures, the actual aspect ratio is not respected. A 3D visualization of the field is also presented in Figure \ref{fig:3d}. 

As shown in Figure \ref{fig:snap_slice}, the moist patches have a conic structures with a larger diameter over the boundary layer and shrinking toward the upper troposphere. Few deep convective towers above 8000 m are observed (Figure \ref{fig:3d}). The higher fluctuations of $h_f$ are located at the top of the boundary layer at about 1000 m and a secondary peak is located around 2000 m. Anomalies tend to vanish above 8000 m. Most of the anomalies of $h_f$ are totally reflected by the total water anomalies, compare Figure \ref{fig:snap_slice}a and b, except in the deepest convective towers where large negative anomalies of $h_L$ can be also found. Small structures like plumes are present in the boundary layer in the form of dry plumes coming from the lower troposphere under the dry region and moist plumes coming from the surface under the moist region. Under deep convective cores in the boundary layer, there are regions with negative vertical velocity (Figure \ref{fig:3d}b) and negative anomalies of $h_L$ which expand toward the surface. We refer to such structures as cold pools.\\

\begin{figure}[t!]
    \centering
    \includegraphics[width=1\linewidth]{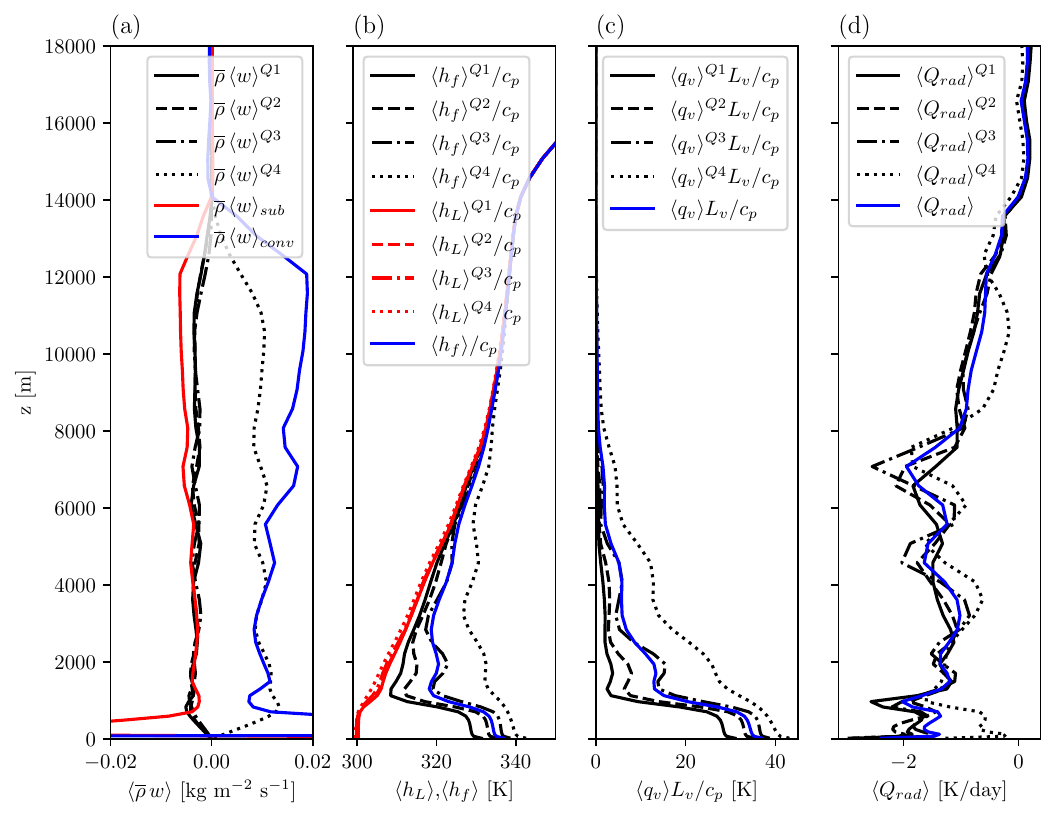}
    \caption{Conditionally averaged profiles of: a) mass flux (red lines indicates the estimate based on the Weak Temperature Gradient (WTG) model), b) liquid ice and frozen moist static energy, c) water vapor, d) radiative heating rate. Profiles are sorted by CRH with Q1 apex indicating the lowest quartile (driest regions) and the Q4 apex indicating the uppermost quartile (moistest regions).}
    \label{fig:cond_prof}
\end{figure}

In order to statistically characterize this pattern, we introduce conditional statistics following the approach by \citet{bretherton2005energy}. All quantities are sorted into four quartiles by the block-averaged Column Relative Humidity (CRH), computed by dividing the simulations domain into blocks with area of 96 $km^2$. The block average (and time average) operator over a certain quartile $Qn$, where $n=(1,2,3,4)$ will be denoted as $\langle\cdot\rangle^{Qn}$. Driest regions correspond to the lowest quartile, while moistest regions to the uppermost quartile.  The resulting average mass flux, liquid ice and frozen moist static energy and water vapor profiles are reported in Figure \ref{fig:cond_prof}.

Moistest regions are inside the cluster where strong convective activity is present. They are not necessarily located on its center, while driest regions are always the furthest regions from the cluster center. On average Figure \ref{fig:cond_prof}a, shows a strong asymmetry on the vertical mass flux between the moistest regions characterized by a strong positive mass flux and all other regions characterized by a negative, although weaker downward mass flux. The mass flux has a bottom heavy structure, linearly increasing (decreasing) in the moistest (driest) regions and reaching a maximum (minimum) value at the top of the boundary layer around 1000 m. Above 2000 m the mass flux is almost constant with height in all regions up to 14000 m where it becomes null. In the following, we will show how the constant value of subsidence in the driest region in the lower free troposphere can be estimated by assuming that the reduced budget (Eq \ref{eq:red_budg_lowt}) is locally valid in such a region. 

The regions where $\langle w \rangle^{Qn}$ is negative will be hereafter referenced as subsiding and all the relative quantities indicated by subscript $sub$, while the regions where $\langle w \rangle^{Qn}$ is positive will be denoted as convective and indicated by subscript $conv$. As highlighted by Figure \ref{fig:cond_prof}a, the quartiles Q1,Q2 and Q3 are part of the subsiding regions with an average CRH lowest than 0.42, while only the uppermost quartile Q4 can be defined as convective, and is characterized by an average CRH of 0.7. Therefore, the subsiding regions occupy an area fraction of $\sigma_{sub}=0.75$ of the total domain while the convective regions only $\sigma_{conv}=0.25$. Taking the average of the mass conservation (Eq. \ref{eq:mass}) for an anelastic system where $\overline{\rho}$ does not change in time and lateral boundary conditions are periodic, we must have that $\langle w \rangle=0$, so we can write
\begin{equation}
  \langle w \rangle_{sub} (1-\sigma_{conv})+ \langle w \rangle_{conv} \sigma_{conv}=0    
  \label{eq:masssub}
\end{equation}
Following previous work by \cite{Robe1996,Agasthya_2025}, in a statistically steady radiative-convective equilibrium, in the dry regions, where convective transport is absent ($\langle w'h_f'\rangle_{sub} \approx 0$), the large-scale radiative cooling must be balanced by subsidence warming. When specializing Equation \ref{eq:red_budg_lowt} for the subsiding regions, identity $\langle w h_f \rangle_{sub} = \langle w' h_f ' \rangle_{sub}$ is no longer valid since $\langle w \rangle_{sub} \neq 0$. Therefore, Equation \ref{eq:red_budg_lowt} is rewritten as:
\begin{equation}
    \frac{\partial \overline{\rho} \langle w h_f \rangle_{sub}}{\partial z}  \approx \overline{\rho}c_{pd}\langle Q_{rad}\rangle_{sub}\\
    \label{eq:red_budg_lowt_sub}
\end{equation}
By assuming that in the lower free troposphere the mass flux is constant with height (Figure \ref{fig:cond_prof}a) and the convective transport is negligible ($\langle w'h_f'\rangle_{sub} \approx 0$) we can write:
\begin{equation}
    \Big\langle w\frac{\partial h_f}{\partial z} \Big\rangle_{sub} \approx \langle w \rangle_{sub} \frac{d\langle h_f \rangle_{sub}}{dz} \approx c_{pd}\langle Q_{rad}\rangle_{sub}
    \label{eq:sub1}
\end{equation}
By considering that in the dry regions $d\langle h_f \rangle_{sub}/dz \approx d\langle h_L \rangle_{sub}/dz$, we can write a scaling for $\langle w \rangle_{sub}$ as following:
\begin{equation}
    \langle w \rangle_{sub} \approx \frac{c_{pd}\langle Q_{rad}\rangle}{d\langle h_L \rangle/dz}
    \label{eq:sub2}
\end{equation}
where we have assumed that $\langle h_L \rangle \approx \langle h_L \rangle_{sub}$ due to the observed small changes of $\langle h_L \rangle$ between dry (subsidence) and moist (convective) regions. The same assumption is made on radiative heating rates, $\langle Q_{rad} \rangle_{sub} \approx \langle Q_{rad} \rangle$. This results in an approximately local radiative–subsidence balance, which can be viewed as a consequence of convective quasi-equilibrium rather than an independent assumption. This is shown to be a good scaling for subsidence velocity in the driest region of the free troposphere (red line in Figure \ref{fig:cond_prof}a). Following Equation \ref{eq:masssub} and by considering our scaling \ref{eq:sub2} for the subsidence velocity, we can also write a scaling for the convective velocity as: 
\begin{equation}
    \langle w \rangle_{conv} \approx \frac{(\sigma_{conv}-1)}{\sigma_{conv}} \frac{c_{pd}\langle Q_{rad}\rangle}{d\langle h_L \rangle/dz}
    \label{eq:sub3}
\end{equation}
By substituting the value $\sigma_{conv}=0.25$ obtained in our simulation, the above scaling leads to the following estimate
\begin{equation}
    \langle w \rangle_{conv} \approx 3 \langle w_{dry}\rangle
\end{equation}
As shown in Figure \ref{fig:cond_prof}a with a blue line, this scaling is a good approximation also for the convective velocity in the moistest regions. 

Figure \ref{fig:cond_prof}b shows the conditional averaged liquid ice static energy, $h_L$, which is almost homogeneous in the near-surface and in the boundary layer up to about 750 m. The maximum difference between moist and dry regions in terms of $h_L$ is found in the boundary layer top, where the moistest regions have $h_L$ values up to 2.5 K smaller than drier regions. This difference reduces significantly above 3000 m, in the mid and upper troposphere, where the values are always lower than 1 K. While homogeneous in terms of $h_L$, the near-surface and the boundary layer exhibit some differences in terms of $h_f$, mostly due to the differences in water vapor content $q_v$ (see Figure \ref{fig:cond_prof}c). In the boundary layer under moistest region the latent heating effect has a value which is about 10 K higher than the driest region. The larger difference between dry and moist region in terms of $h_f$ it is of about 30 K and it is always found in the boundary layer top. Finally it is important to notice that the gradient of $h_f$ in the vertical direction is much larger in the driest regions than in the moistest ones, especially at the boundary layer top. By looking at the conditionally averaged radiative cooling rate $Q_r$ in Figure \ref{fig:cond_prof}d, this large gradient coincides with a negative peak of radiative cooling and a corresponding peak of shallow clouds (not shown) in the Q3 moist quartile.

\begin{figure}[t]
    \centering
    \includegraphics[width=0.7\linewidth]{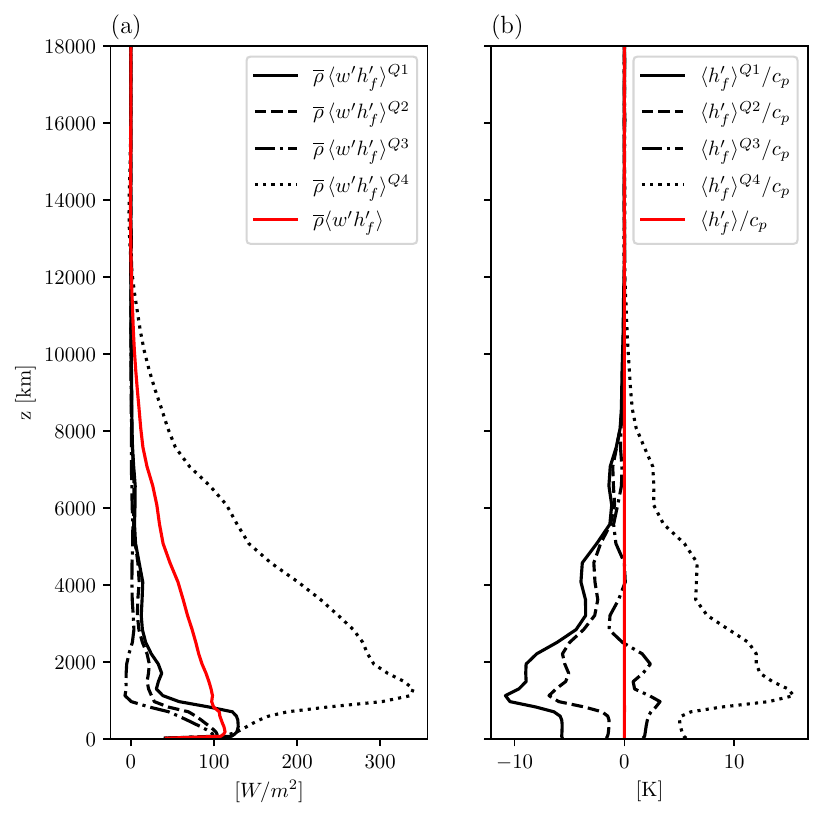}
    \caption{Conditionally averaged profiles of: a) vertical convective transport of frozen moist static energy $\overline{\rho} \langle w'h_f' \rangle$; b) frozen moist static energy anomaly $\langle h_f' \rangle/c_p$. Profiles are sorted by CRH with Q1 apex indicating the lowest quartile (driest regions) and the Q4 apex indicating the uppermost quartile (moistest regions).}
    \label{fig:cond_prof_trans}
\end{figure}

The scalings for the subsidence and convective velocities, \ref{eq:sub2} and \ref{eq:sub3}, hold well above the boundary layer top region, and are based on the assumption of negligible convective transport in dry regions, $\langle w'h_f'\rangle_{sub} \approx 0$. As shown in Figure \ref{fig:cond_prof_trans}a, this assumption is valid in the subsiding regions (Q1,Q2,Q3), confirming our hypothesis. On the other hand, the transport becomes significant in all regions below the boundary layer top region. In the moistest region the peak value of convective transport is reached above the boundary layer top at about 1300 m and it is approximately 3 times larger than the surface total frozen moist static energy fluxes. Regarding the conditional anomaly of $h_f$, the intermediate moist regions (Q3) show an almost null anomaly, while the fluctuations in the moistest regions (Q4) are two times larger than the ones in the driest regions (Q2) and of the order of 10 K. 

In conclusion, the boundary layer top is a region of strong inhomogeneities of $h_f$ both in the vertical and in the horizontal direction between dry and moist regions. The strongest inhomogeneity is found in the driest regions and in the intermediate moist region. In the driest region, this is especially linked to the mean vertical structure of water vapor and the very dry subsiding air above a relatively moist boundary layer. 
Above the boundary layer top, the subsiding velocity in the driest regions free of convection is found to scale with a good approximation to the ratio of the large-scale average radiative cooling, $c_{pd} \langle Q_{rad} \rangle$, and the large-scale stability, $d\langle h_L \rangle/dz$, confirming that the dynamics outside the convective regions is governed by radiative balance \cite{Agasthya_2025,Robe1996}. When the fraction of area occupied by convective regions, $\sigma_{conv}$ is given, such scaling leads to a correct estimate also of the convective velocity. Alternatively, as also demonstrated by \cite{Agasthya_2025}, the average convective velocity, $\langle w\rangle_{conv}$, can be used as velocity scale to correctly adimensionalize the problem. Another conclusion is that in convective regions, the surface layer, boundary layer, and the various layers of the troposphere are “well mixed”; in fact, the gradients of the profiles of various average quantities are more homogeneous in the uppermost quartile Q4. In contrast, in dry regions (Q1, Q2, and Q3 quartiles), average gradients form in the boundary layer that highlight separate dynamics above and below the boundary layer. This separation between the boundary layer and the free troposphere in dry regions is also evident when examining the turbulent convective fluxes in Figure \ref{fig:cond_prof_trans}a, which are zero above the boundary layer.

\section{Atmospheric mixing and organization: energy variance budget}
\label{sec:variance}

\begin{figure}[t]
    \centering
    \includegraphics[width=0.8\linewidth]{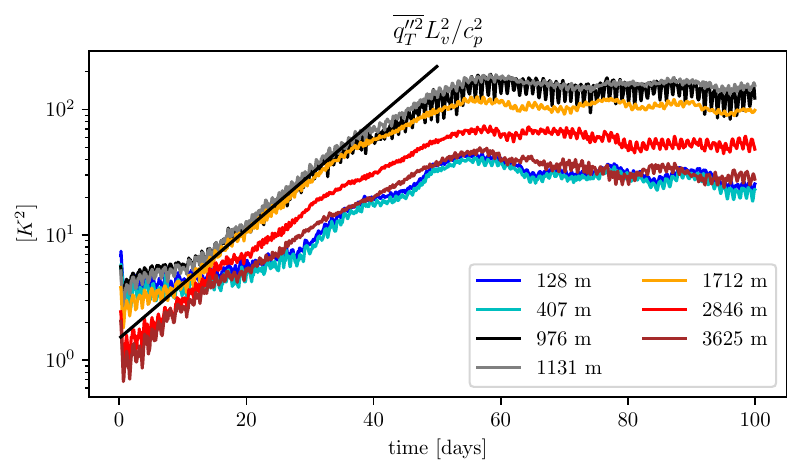}
    \caption{Time evolution of total water variance. Different colors show different model levels. Black line shows a exponential fit of the variance increase with an e-folding time of the order of 10 days.}
    \label{fig:qtvar_evo}
\end{figure}

\begin{figure}[t]
    \centering
    \includegraphics{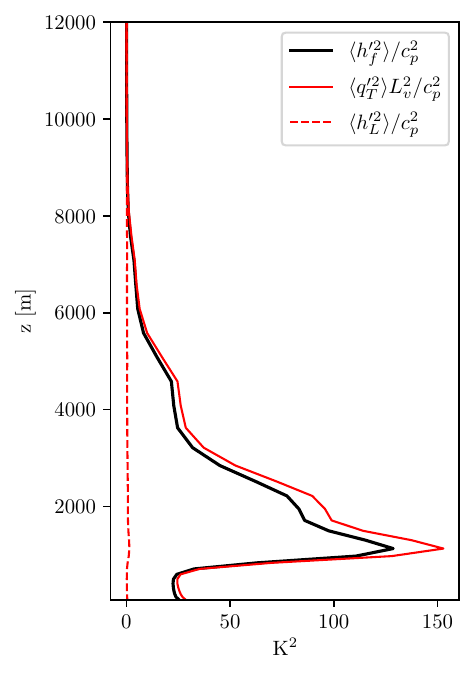}
    \caption{Vertical profile of energy variance: $\langle h_f'^2 \rangle$ is shown in black line, the $\langle q_T'^2 \rangle$ variance is shown in solid red line and the $\langle h_L'^2 \rangle$ variance is shown in dashed red line.}
    \label{fig:hf2_vert}
\end{figure}

The transition from a random cluster to an aggregated state with a very dry and very moist region denotes an increase in the energy variance of the system $\langle h^{\,\prime\,2}_f \rangle$. Figure \ref{fig:qtvar_evo} shows the evolution in time of the total water variance at different atmospheric levels. The total water variance represents most of the frozen moist static energy variance in the domain (see Figure \ref{fig:snap_slice} and following discussion on Figure \ref{fig:hf2_vert}), therefore it is shown here for simplicity. At the initial steps of the simulation, up to 10 days (240 hours), the variance is almost constant below 2000 m, while it is exponentially increasing on the above layers. This increase is associated to the formation of deep convective cores and also persistent dry patches above 2000 m. Only after 10 days, the variance starts increasing exponentially also in the boundary layer and up to 2000 m. The black solid line indicates an exponential fit with an e-folding time of the order of 10 days. Until 40 days (960 hours) the variance increases exponentially in the boundary layer with the same rate. During this period, dry patches are penetrating in the boundary layer and expanding. Between 40 and 60 days, the growth rate decreases, and then it stabilizes reaching an asymptotic equilibrium after 60 days. This period between dry patch expansion and the single cluster steady state is associated with the merging of different dry patches.

The non-homogeneous vertical distribution of energy variance $\langle h_f^{'2} \rangle$, evaluated in the final equilibrium state, is reported in Figure \ref{fig:hf2_vert}. The peak of energy variance, $\langle (h_f^{'2})\rangle_{max}$ is reached at the boundary layer top, around 1100 m. From this peak, the energy variance decreases by moving towards the upper troposphere and towards the surface layer. Below 500 m and in the surface layer the variance is almost constant with height and has values of one order of magnitude smaller than the boundary layer top. Almost all the energy variance can be shown to be related to fluctuations of water vapor (solid red line in Figure \ref{fig:hf2_vert}), with small fluctuations of liquid-ice static energy (solid red line in Figure \ref{fig:hf2_vert}). 

In order to study how such energy variance $\langle h^{\,\prime\,2}_f \rangle$ is produced and transported across the domain, we analyze here its evolution equation (full derivation in Appendix A):
\begin{align}
    \frac{ \partial \rhobar \langle h_f'^2 \rangle}{\partial t} &+ \frac{\partial \rhobar \langle h_f'^2 \rangle U_j}{\partial x_j} + \frac{\partial \rhobar \langle h_f'^2 u_j' \rangle}{\partial x_j} + 2\overline{\rho}\langle u_j' h_f' \rangle \frac{\partial H}{\partial x_j} =\\
    &=-2 L_f  \langle h_f'\frac{\partial P_f'}{\partial z} \rangle - 2 \frac{\partial \langle \tau_{h_f,j}' h_f' \rangle}{\partial x_j} -2\langle\epsilon^{h}_{sgs} \rangle + 2\rhobar c_{pd} \langle h_f' Q_{rad}' \rangle
    \label{eq:var}
\end{align}
where $\epsilon^{h}_{sgs}=-\tau_{h_f,j}'( \partial h_f'/\partial x_j)$ is the flux of variance of frozen moist static energy from/to subgrid scale, that is the actual dissipation, $U_j \equiv \langle u_j \rangle$ and $H \equiv \langle h_f\rangle$. Due to the statistical symmetries of the deep convection radiative-convective equilibrium, the energy variance budget simplifies as:
\begin{align}
    \underbrace{\frac{d \rhobar \langle w' h_f'^2\rangle}{dz}}_{ADV}&+ \underbrace{2\frac{d \langle\tau_{h_f,z}'h_f'\rangle}{dz}}_{DIFF}= \nonumber \\
    &=\underbrace{-2\overline{\rho}\langle w' h_f'\rangle\dv{H}{z}}_{GRAD} \underbrace{-2 L_f \langle h_f'\frac{d P_f'}{dz}\rangle}_{PREC}-\underbrace{2\langle\epsilon^{h}_{sgs}\rangle}_{DISSIP} + \underbrace{2\rhobar c_{pd} \langle h_f' Q_{rad}'\rangle}_{RAD}
    \label{eq:varsymm_main}
\end{align}
Equation \ref{eq:varsymm_main} can be rewritten also in a conservative form as:
\begin{equation}
    \frac{d \Phi_{h_f^{\prime 2}}}{dz}= \mathcal{S}
    \label{eq:var_flux}
\end{equation}
thus highlighting that the fluxes of frozen moist static energy are statistically occurring only in the vertical direction and are driven by local source/sink mechanisms. In the above equation 
\begin{equation}
    \Phi_{h_f^{\prime 2}}=\rhobar \langle w' h_f'^2\rangle+2\langle\tau_{h_f,z}'h_f'\rangle
\end{equation}
is the total flux of FMSE variance, that is the sum of the vertical advective and vertical diffusive flux of FMSE variance. The source/sink of such flux,
\begin{equation}
    \mathcal{S} = \underbrace{-2\overline{\rho}\langle w' h_f'\rangle\dv{H}{z}}_{GRAD} \underbrace{-2 L_f\langle h_f'\frac{d P_f'}{dz}\rangle}_{PREC}-\underbrace{2\langle\epsilon^{h}_{sgs}\rangle}_{DISSIP} + \underbrace{2\rhobar c_{pd} \langle h_f' Q_{rad}'\rangle}_{RAD}
\end{equation}
is due to the vertical mean gradient of $h_f$ (GRAD, gradient term), the divergence of frozen precipitation fluxes (PREC, precipitation term), dissipation (DISSIP) and radiation (RAD).

\subsection{Source and sink of energy variance}

\begin{figure}[t]
    \centering
    \includegraphics[width=1\linewidth]{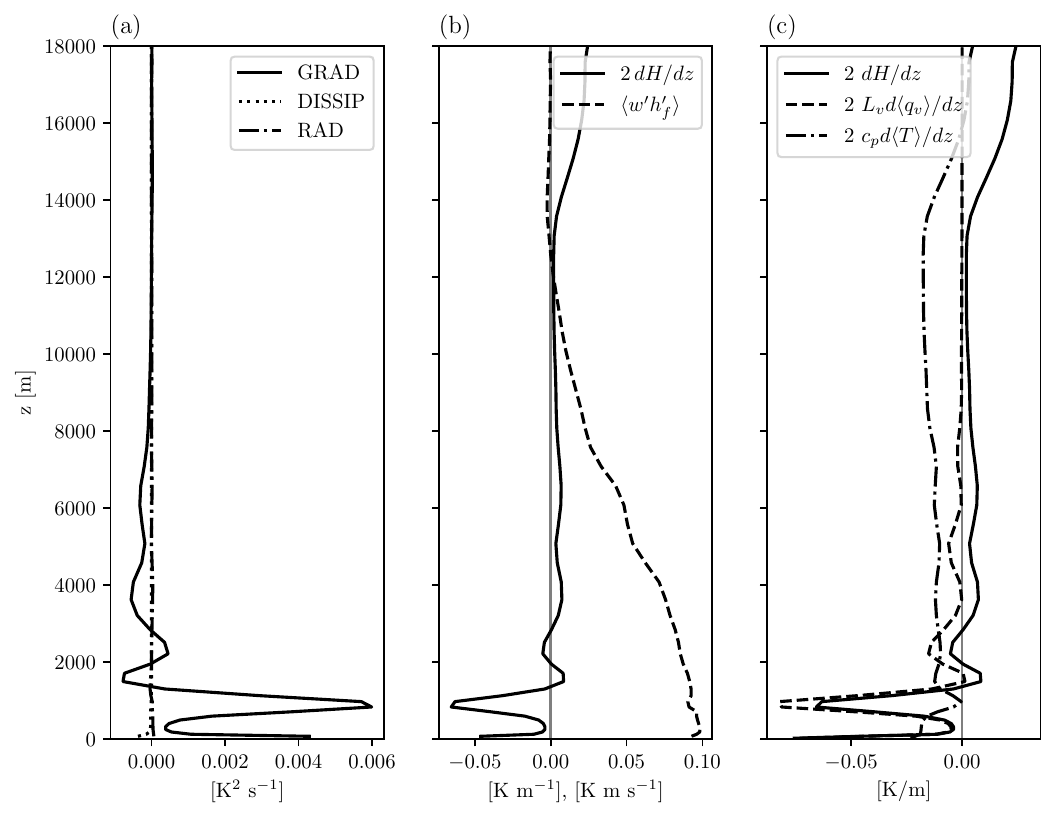}
    \caption{a) Fluxes of FMSE variance $\Phi_{h_f^{\prime 2}}$, b) Source/sink term of fluxes of FMSE variance, c) Components of the gradient source/sink term, d) Components of the large-scale gradient of $h_f$.}
    \label{fig:prod_terms}
\end{figure}

As shown in Figure \ref{fig:prod_terms}a, the only relevant source of energy variance is related to the mean gradient $dH/dz$ (GRAD term). Radiation is negligible throughout the atmosphere, while dissipation is a non-negligible sink only in the surface layer. Starting from the near-surface layer, the energy variance source by mean gradient exhibits a first peak around 250 m followed by a rapid decrease in the boundary layer region from 250 to 500 m. It then increases up to reach its highest value in the boundary layer top region around 800 m just below the peak of energy variance around 1100 m. By further moving upward, the source of FMSE variance from the gradient term decreases reaching values that oscillate around 0 becoming first a sink between 1300 and 2000 m in the shallow layer clouds region, then a source between 2000 m and 2500 m in the mid troposphere and then again a sink until 8000 m in the upper troposphere, where it approaches negligible values.

The source of energy variance by the gradient term is given by the product of the vertical gradient $dH/dz$ with the vertical fluxes of mean frozen moist static energy $\langle \overline{\rho} w'h_f' \rangle$. As already shown in Figure \ref{fig:mean_flux} when analyzing the vertical transport of $H$ and again shown here in Figure \ref{fig:prod_terms}b for readability, the latter is always positive across the vertical. Hence, the shape of the source of energy variance by the gradient term strictly follows the behaviour of vertical gradient of mean frozen moist static energy $dH/dz$. By decomposing the gradient $dH/dz$ in its contributions, Figure \ref{fig:prod_terms}c shows that is almost entirely determined by the vertical gradient of mean water vapor, $d\langle q_v \rangle/dz$. In particular, the first peak of production in the near-surface layer of the gradient term is the signature of the strong input of moisture from the surface. Another consequence of the fixed sign of the vertical fluxes $\langle \overline{\rho} w'h_f' \rangle$ is that, on average, a positive fluctuation of energy, $h'> 0$, is statistically associated to an upward motion, $w'> 0$, while a negative fluctuation to a downward motion, independently on the sign of the mean gradient $dH/dz$. This observation suggests that the classical Boussinesq assumption of turbulent fluxes that go down the mean gradient (eddy viscosity and turbulent diffusivity assumptions) actually fails in describing the mixing of frozen moist static energy. This fail also indicates that vertical fluxes are not associated to local mixing phenomena, rather to long distance interactions \citep{cimarelli2019negative}. Hence, we conjecture that the source of energy variance is mainly associated to the large-scale circulation between moist and dry regions. 

Overall, the role of the frozen moist static energy gradient is to modulate the intensity and also the sign of the source term, with the radiative term having negligible influence. By assuming that the source of energy variance by the gradient term is driven by the state of large-scale aggregation of moist and dry regions, this observation suggests that radiation is not a physical process directly contributing to the aggregated state of the atmosphere. Radiation contributes to the energy variance production only indirectly by determining the shape of the FMSE gradient, $dH/dz$, as already shown in Figure \ref{fig:mean_flux}c when analysing the budget of mean FMSE, Equation \ref{eq:hf_budg_mean}. 

It is relevant now to address also the horizontal inhomogeneity of the energy variance source.
The two components of the gradient term, $dH/dz$ and $\rho \langle w'h_f'\rangle$, are found to be almost uniform in the dry and the moist region as already shown in Figures \ref{fig:cond_prof} and \ref{fig:cond_prof_trans}. Therefore, the mechanism of production of energy variance in the near surface layer seems to be almost the same between dry and moist region. In the boundary layer, production tends to decrease due the low values of $dH/dz$ attained there. The second large production peak is on the boundary layer top region, where the strongest gradients are present. In the dry regions, the lower values of $\langle w'h_f'\rangle$ are compensated by larger mean gradient $dH/dz$ caused by the contrast between the relatively moist boundary layer air and the dry subsiding air from above. On the other hand in the moist regions, the mean gradient is weak but the turbulent fluxes are very large causing the production to be substantial. Two different key mechanisms generating energy variance are at play above the boundary layer top region which are essential to the maintenance of the self-aggregation. 
The shallow clouds layer is a dissipative region, where the source of the gradient term becomes negative, actually a sink. In this region, a strong asymmetry of the vertical turbulent fluxes is present between the dry subsiding and moist convective part of the domain. While in the dry regions such fluxes are almost null, in the moist regions they reach their maximum values around 1300 m (Figure \ref{fig:cond_prof_trans}a). Regarding the gradient $dH/dz$, this is slightly positive in the dry regions, while negative and almost null in the moist regions. This change of sign of $dH/dz$ in the driest regions is mainly related to water vapor (Figure \ref{fig:cond_prof}b and Figure \ref{fig:cond_prof}c). Above the shallow cloud layer in the mid-troposphere another small region of production is present, that must be related to moist regions, since the turbulent fluxes in the dry regions becomes negligible. Finally all energy mostly produced in the near-surface layer and boundary layer is dissipated throughout the upper troposphere where the gradient of $H$ start to be positive and the variance fluxes start decreasing up to reaching the convective scale height of 12000 m. 

\subsection{Fluxes of energy variance}

\begin{figure}[t]
    \centering
    \includegraphics[width=1\linewidth]{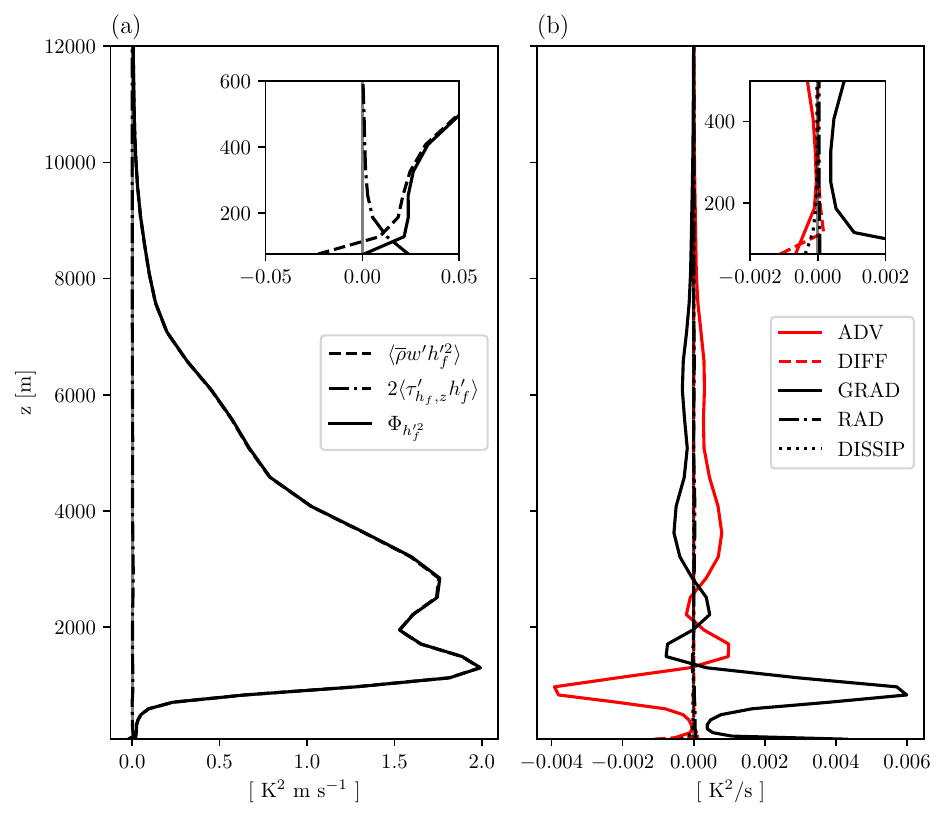}
    \caption{a) Fluxes of FMSE variance $\Phi_{h_f^{\prime 2}}$, b) Terms of energy variance budget as reported in Equation \ref{eq:varsymm_main}. All the terms have been rescaled by the factor $(\overline{\rho} c_p^2)$.}
    \label{fig:hf2_flux_budget}
\end{figure}

The net excess of energy variance produced in the source regions is transported towards the sink regions where is eventually dissipated by a field of fluxes $\Phi_{h_f^{\prime 2}}$ described by Equation \ref{eq:var_flux}. As shown in Figure \ref{fig:hf2_flux_budget}a, the fluxes of FMSE variance are always positive, $\Phi_{h_f^{\prime 2}} \ge 0$, thus showing an overall upward transport. Starting the analysis from the ground, it is possible to observe that the flux attains vanishingly small values in the surface and boundary layer. Hence, the energy variance produced in this layer by the first peak of activity of the gradient term remains confined to this region, thus highlighting a decoupling of the energy variance dynamics of this layer with respect to those occurring at higher elevations. Indeed, the vertical flux starts to increase at the boundary layer top, fed by the peak activity of the source term in this region. Hence, a net upward flux traverses all the atmospheric layer from the boundary layer top to the troposphere thus highlighting a strongly coupled dynamics of these layers. Two peak values of vertical flux occur at about 1300 m and 2600 m. The former, higher one is located slightly above the peak of the source at 800 m and of the energy variance at 1100 m. From these peak values, the flux gradually decreases by moving toward the upper troposphere. As shown in Figure \ref{fig:prod_terms}a, almost all the flux of energy variance is related to vertical advection. The only exception is the near-surface layer, below 250 m, where diffusive fluxes become relevant. Interestingly, in the near-surface layer the advective term becomes negative thus representing a downward flux of energy variance. This negative value is however balanced by the positive value of diffusive fluxes resulting in a net zero transport at the surface.

\subsection{Overall mixing and structure of energy variance}

The overall view provided by the energy variance equation in the conservative form, is that of a mixing of the surface and boundary layer that is decoupled from that occurring from the boundary layer top to the upper troposphere. In particular, in the surface and boundary layer the spatial flux is almost negligible and the local source of energy variance by the gradient term remains confined and dissipated there. This fully mixed equilibrium leads to an almost homogeneous distribution of energy variance within these layers. On the other hand, the layers from the boundary layer top to the upper troposphere are fully coupled being traversed by an upward advective flux of energy variance. A peak of energy variance source occurs at the boundary layer top and is performed by the gradient term and is related to the strong vertical inhomogeneity of mean water vapor content occurring in this region. This peak of source sustains energy variance locally (the peak of energy variance is located in this layer) and feeds the upward advective flux that eventually is responsible in sustaining the levels of energy variance of the troposphere that actually acts as a sink. Note that from the peak value at the boundary layer top, the source region of energy variance extends up to the lower troposphere. The only exception is given by the shallow clouds layer that acts as an intermediate layer as energy variance sink in between an overall source region. The sink is related to the increase of $H$ with height in this layer, mainly caused by the increase of $\langle q_v \rangle$, since the other term $\langle q_c \rangle$ contributing to $H$ actually decreases in this layer as shown in Figure \ref{fig:mean_prof}c. This increase is mainly found in the driest and intermediate regions as shown in Figure \ref{fig:cond_prof}b and \ref{fig:cond_prof}c, since $H$ is constant in the moistest region in this layer. By looking at the budget of the average non precipitating water (thus mainly related to water vapor, not shown), two mechanisms act as sources in this region: rain evaporation and vertical advection of water vapor. Rain evaporation can occur in the intermediate moist regions where cloud and rainfall are found. However, in this intermediate moist region (Q3), as we have shown in Figure \ref{fig:cond_prof_trans}a, the conditionally averaged fluxes are almost null or negative, $\langle w'h_f'\rangle \leq 0$, therefore causing the shallow cloud layer in the intermediate moist regions being a locally positive (or negligible) source region. In this case turbulent fluxes of $h_f$ (maybe also linked to rain evaporation in congestus clouds) are going down the mean gradient thus sustaining the variance (moistening an already moist region). On the other hand in the driest regions (Q1) the turbulent fluxes are very high close to the surface and tend to sustain moist plumes from the boundary layer which penetrates the boundary layer top region and tend to moisten the shallow cloud layer (slightly positive fluxes $\langle w'h_f'\rangle <0$ at around 2000 m in Figure \ref{fig:cond_prof_trans}a). Therefore in this driest region the advection of water vapor in the boundary layer top may be responsible for the energy variance sink, by moistening a drier than average region. In the intermediate dry region (Q2) both rain evaporation and vertical advection maybe positively contributing to the sink layer of variance.\\
A reduced description of this complex scenario is provided by the budget of energy variance (Equation \ref{eq:varsymm_main}). As shown in Figure \ref{fig:hf2_flux_budget}b, the atmospheric layers from the ground up to the lower troposphere are net sources of energy variance with the only exception of the shallow clouds layer. In this regions, the advective transport term is negative thus draining the source of energy variance to feed other atmospheric layers. Indeed, the troposphere is a net sink and energy variance there is uniquely sustained by the advective transport term that is positive thus releasing the energy variance drained at lower atmospheric layers. We can now provide a reduce relation able to describe, in a simplified form, the overall budget of energy variance. As shown in Figure \ref{fig:hf2_flux_budget}b, the contributions of diffusion and sub-grid scale dissipation are always negligible except to the near-surface layer. Hence, the budget of energy variance simply reduces to a balance between the source provided by the mean gradient of FMSE and the vertical transport performed by advection,
\begin{equation}
    \frac{d \rhobar \langle w' h_f'^2\rangle}{dz} \approx -2\overline{\rho}\langle w' h_f'\rangle \frac{dH}{dz}
    \label{eq:var_budget_red}
\end{equation}
This reduced-order model is valid throughout the atmosphere except in the very-near surface layer where transport mechanisms related to diffusion and sink phenomena related to subgrid dissipation become relevant.

\clearpage

\section{Discussion and conclusions}
\label{sec:conclusions}

This work represents the first part of a research activity that regards the characterization of mean quantities, conditional statistics and energy variance budget in the final stage of convective self-aggregation, when a single steady cluster of convection has formed. The energy variance budget analysis is conducted over all the atmospheric layers, in a similar way to the Vertically-Resolved framework introduced by \cite{yao2022vertically}. However, instead of calculating the adiabatic production as a residual and interpreting it as the sum of horizontal and vertical advection of FMSE, we directly compute the two different terms contributing to the adiabatic part of the energy variance budget \citep{khairoutdinov2002similarity}: the vertical advection of energy variance, $ADV=d \rhobar \; \langle w' h_f^{\prime 2} \rangle / dz$ and the gradient term, $GRAD=-2 \rhobar \langle w' h'_f\rangle dH/dz$. Above the near-surface layer (250 m), we found that in the final steady state of convective self-aggregation this two terms closely balance each other. Instead near the surface layer, transport mechanisms related to diffusion and sink phenomena related to subgrid dissipation become relevant. 

Whether the gradient term represents a source or a sink of energy variance depends directly on the sign of the mean FMSE vertical gradient, $dH/dz$, since the vertical fluxes of FMSE anomalies are found to be always positive, $\langle w'h'_f \rangle>0$. Up to the boundary layer top at about 1300 m, the gradient term is always positive. Physically, in this region, we can interpret the gradient term as the adiabatic generation of local FMSE variance through the coupling between the convective overturning circulation, $\langle w' h'_f \rangle$, and the mean vertical stratification, $dH/dz$. Below 1300 m above ground, the background state FMSE decreases sharply with altitude ($dH/dz < 0$). Consequently, the positive covariance $\langle w' h_f' \rangle > 0$ across the domain systematically amplifies horizontal anomalies. As confirmed by the conditional analysis, in moist regions, ascending motions ($w'>0$) advect high-FMSE air upward from the surface layer, further strengthening the positive local anomaly ($h'_f > 0$). Conversely, in dry regions, subsiding motions ($w'<0$) forces the intrusion of dry, low-FMSE mid-tropospheric air into the boundary layer, intensifying the negative local anomaly ($h'_f < 0$). This dual process acts as a powerful thermodynamic feedback that continuously dries the dry environments and moistens the wet environments. This mechanism effectively converts the vertical gradient of the base state into horizontal mesoscale variance, confirming the primary role of the boundary layer and low-level circulations in convective self-aggregation \cite{Muller2015whatfavors,yang2018boundary,yao2022vertically}.

The relevance of the gradient term may remain hidden when considered together with vertical advection as a whole adiabatic production term. The close balance between such terms may enlarge the relative importance of radiative feedbacks in the energy variance budget, especially at the final stage of convective aggregation as found in \cite{yao2022vertically}. In our study we demonstrate that in the maintenance phase of self-aggregation, radiative heating anomalies (or differential radiative cooling) are negligible. Radiation contributes to the variance production only indirectly through the definition of the mean FMSE gradient, $dH/dz$. In particular gradient term shows two large positive peaks of energy variance production at the surface and at about 1000 m, respectively. Such peaks directly corresponds to the highest peaks of the mean profiles of humidity, $d\langle q_v \rangle/dz$, which we have shown to correspond to the largest peaks of average radiative cooling profiles, $\langle Q_{rad} \rangle$, characterizing the driest and intermediate regions of the domain. Obviously, at the surface such strong gradients are caused also by surface fluxes. The results obtained within this variance framework for the boundary layer in deep-convective self-aggregation are also in agreement with the radiatively cooled convective boundary layers simulated by \cite{de2004large}, where the production of virtual potential temperature variance has a maximum near the inversion layer. Moreover, the relevance of the domain-mean FMSE profile and its gradient could be used to explain the different response of self-aggregation on radiative feedback over different SST \citep{arnold2013enhanced,yao2023convective}. The indirect role of radiative cooling as found in this work seems to be in agreement with previous recent studies on shallow cumulus aggregation where both simulations \citep{Bretherton2017mesoscaleagg,janssens2023nonprecipitating,Janssens2024shallow} and observations \citep{george2023widespread} show how shallow mesoscale circulations are not directly driven by radiative cooling.

Within the shallow cloud layer, between 1300 and 2000 m, the gradient term becomes a sink of energy variance, mainly due to a slightly positive gradient, $dH/dz$, linked to a local increase in water vapor in the driest and intermediate regions of the domain. Above the shallow cloud layer, in the mid-troposphere another small region of production is present that is related to the moist regions, since convective fluxes, $\langle w'h'_f \rangle$, are almost null over the dry regions at such heights. Finally, all energy produced mostly at the surface and boundary layer is dissipated throughout the upper troposphere, where the gradient term is always a sink. 

\subsection{Follow-up}

Diabatic sources such as radiative cooling and surface fluxes fundamentally act as local thermodynamic forcing functions, lacking intrinsic spatial transport capabilities. The translation of these local thermodynamic anomalies into remote dynamical responses has to be mediated by the overturning circulations and convection, which redistribute FMSE across the space and across different scales. Other important processes in the boundary layer could be associated to the evaporation of rainfall and the formation of cold pools. As hypothesized by \cite{thomas2021toward} cold pools are responsible for organizing small randomly distrubuted shallow cumulus clouds into larger clusters or mesoscale arcs, determining an increase of total water variance as shown by other studies \citep{Bretherton2017mesoscaleagg,schemann2017budget}. This multi-scale dynamical feature of circulation in convective self-aggregation supports the adoption of a scale-by-scale formalism able to describe it. This new framework and results on how locally generated energy variance cascades through the system, will be reported in the second part of our research activity \citep{silvestri2026gye_part2}.

\acknowledgments
L.S. and P.B.C was financed by the PR FESR 2021-2027 project - Action 2.4.1 “Prevention of Natural Hazards” - RIMU-CLIMA Project “Integrated Umbrian Meteorological Network and Tools for Climate Analysis in Umbria” (CUP:  I62B25000210006) Italian Ministry of University and Research (MUR). Moreover, this research has been also carried out as part of the PRIN 2022 PNRR Project P20229KW2R - SEAPLANE - "Simulation and modelling of interface fluxes in wind-wave flows for an improved climate science", CUP E53D23017010001, funded by the National Recovery and Resilience Plan (PNRR), Italy, Mission 04 Component 2 Investment 1.1 – NextGenerationEU. We thank Dr. Gabriele Boga for useful discussion on this work, especially for realizing \ref{fig:3d}.

%
%
\datastatement
The SAM model code is freely available at \url{http://rossby.msrc.sunysb.edu/SAM.html}. The main simulation outputs and related statistics are freely available in \cite{silvestri_2026_24336231}.








%



\appendix[A]

\appendixtitle{Derivation of FMSE variance equation}
In the following, we derive the FMSE variance equation. First, we divide Eq.\ref{eq:hf} by $\overline{\rho}$ to get:
\begin{equation}
    \frac{\partial h_f}{\partial t} + \frac{1}{\overline{\rho}}\frac{\partial \overline{\rho}u_j h_f}{\partial x_j}=-\frac{1}{\overline{\rho}}\frac{\partial L_f P_f}{\partial z} - \frac{1}{\overline{\rho}} \frac{\partial \tau_{h_f,j}}{\partial x_j} + c_{pd}Q_{rad}
    \label{eq:hrho}
\end{equation}
In order to derive the equation for the variance, we decompose $h_f$ in its mean and fluctuating part as $h_f= \langle h_f \rangle + h_f'= H+h_f'$, where the operator $ \langle \cdot \rangle$ indicates the average operator and $H \equiv  \langle h_f \rangle$. By applying the average operator to Eq. \ref{eq:hrho}, we can write the governing equation for the mean FMSE:
\begin{equation}
    \frac{\partial H}{\partial t}+\frac{1}{\overline{\rho}}\frac{\partial \overline{\rho} \langle u_j h_f \rangle}{\partial x_j} = -\frac{L_f }{\overline{\rho}}\frac{\partial \langle P_f \rangle}{\partial z} -  \frac{1}{\overline{\rho}}\frac{\partial \langle \tau_{h_f,j} \rangle}{\partial x_j} + c_{pd} \langle Q_{rad} \rangle
    \label{eq:meanhrho}
\end{equation}
By subtracting the mean equation \ref{eq:meanhrho} to the total one \ref{eq:hrho} we get:
\begin{equation}
    \frac{\partial h_f'}{\partial t} + \frac{1}{\overline{\rho}}\frac{\partial}{\partial x_j}\big (\overline{\rho}u_j h_f-\overline{\rho} \langle u_j h_f \rangle \big )=-\frac{L_f }{\overline{\rho}} \frac{\partial P_f'}{\partial z} - \frac{1}{\overline{\rho}}\frac{\partial \tau_{h_f,j}'}{\partial x_j} + c_{pd}Q_{rad}'
    \label{eq:hp}
\end{equation}
that is the equation for the anomaly of the frozen moist static energy. By decomposing the velocity and energy fields in their mean and fluctuating part, we can write the second term inside the brackets as:
\begin{align}
    \overline{\rho} \big (u_j h_f- \langle u_j h_f \rangle \big )&= \overline{\rho}\big (\xcancel{U_j H}+ U_j h_f'+u'_j H+u'_j h_f'-\xcancel{ \langle U_j H \rangle}-\cancelto{0}{ \langle u'_j H \rangle} - \cancelto{0}{ \langle U_j h_f' \rangle}- \langle u'_j h_f' \rangle \big ) \nonumber \\
     &=\overline{\rho} \big (U_j h_f'+u'_j H+u'_j h_f'- \langle u'_j h_f' \rangle \big )
\end{align}
Then the equation for the fluctuating energy becomes:
\begin{align}
    \frac{\partial h_f'}{\partial t} &+ \irhobar \frac{\partial \rhobar U_j h_f'}{\partial x_j} +\irhobar \frac{\partial \rhobar u'_j H}{\partial x_j} + \irhobar \frac{\partial \rhobar u'_j h_f'}{\partial x_j} -\irhobar \frac{\partial \rhobar  \langle u'_j h_f' \rangle}{\partial x_j}= \nonumber \\
    &-\frac{L_f }{\overline{\rho}} \frac{\partial P_f'}{\partial z}- \frac{1}{\overline{\rho}} \frac{\partial \tau_{h_f,j}'}{\partial x_j} + c_{pd}Q_{rad}'
    \label{eq:hp1}
\end{align}
We multiply such equation by $2\rhobar h_f'$ and considering also the continuity equation $\pdv*{(\rhobar u'_i)}{x_i}=0$, we can write:
\begin{align}
    \frac{\partial \rhobar h_f'^2}{\partial t} &+ \frac{\partial \rhobar U_j h_f'^2}{\partial x_j} +2\overline{\rho} u'_jh_f' \frac{\partial H}{\partial x_j} + \frac{\partial \rhobar u'_j h_f'^2}{\partial x_j} -2h_f' \frac{\partial \rhobar  \langle u'_j h_f' \rangle}{\partial x_j}= \nonumber \\
    &-2 L_f h_f'\frac{\partial P_f'}{\partial z}- 2h_f' \frac{\partial \tau_{h_f,j}'}{\partial x_j} + 2\rhobar  c_{pd} h_f' Q_{rad}'
    \label{eq:hp2}
\end{align}
Rearranging the subgrid diffusion term, we have:
\begin{align}
    \frac{\partial \rhobar h_f'^2}{\partial t} &+ \frac{\partial \rhobar U_j h_f'^2}{\partial x_j} +2\overline{\rho} u'_jh_f' \frac{\partial H}{\partial x_j} + \frac{\partial \rhobar u'_j h_f'^2}{\partial x_j} -2h_f' \frac{\partial \rhobar  \langle u'_j h_f' \rangle}{\partial x_j}=\nonumber \\
    &-2 L_f h_f'\frac{\partial P_f'}{\partial z} - 2\frac{\partial \tau_{h_f,j}'h_f'}{\partial x_j} +2\tau_{h_f,j}'\frac{\partial h_f'}{\partial x_j} + 2\rhobar c_{pd} h_f' Q_{rad}'
    \label{eq:hp3} 
\end{align}
By applying the average operator, we finally get the final form of the equation for the FMSE variance:
\begin{align}
    \frac{\partial \rhobar  \langle h_f'^2 \rangle}{\partial t} &+ \frac{\partial \rhobar U_j  \langle h_f'^2 \rangle}{\partial x_j} + \frac{\partial \rhobar  \langle u'_j h_f'^2 \rangle}{\partial x_j} +2\overline{\rho} \langle u'_jh_f' \rangle \frac{\partial H}{\partial x_j} = \nonumber \\
    &-2 L_f  \langle h_f'\frac{\partial P_f'}{\partial z} \rangle - 2\frac{\partial \langle \tau_{h_f,j}'h' \rangle}{\partial x_j} -2 \langle \epsilon^{h}_{sgs} \rangle + 2\rhobar c_{pd} \langle h_f' Q_{rad}' \rangle
    \label{eq:varapp}
\end{align}
where $\epsilon^{h}_{sgs}=-\tau_{h_f,j}'( \partial h_f'/ \partial x_j)$ is the flux of variance of frozen moist static energy from/to subgrid scale, that is the actual dissipation. For the specific settings of the RCE simulation of the present work, some statistical symmetries emerge (statistical homogeneity in time and in the horizontal periodic directions) that allow us to simply the equation as:
\begin{align}
    \frac{d \rhobar \langle w' h_f'^2\rangle}{dz}&+ 2\frac{d \langle\tau_{h_f,z}'h_f'\rangle}{dz}= \nonumber \\
    &=-2\overline{\rho}\langle w' h_f'\rangle \frac{dH}{dz} -2 L_f\langle h_f'\frac{d P_f'}{dz}\rangle -2\langle\epsilon^{h}_{sgs}\rangle + 2\rhobar c_{pd} \langle h_f' Q_{rad}'\rangle
    \label{eq:varsymm}
\end{align}
where all the transport terms have been moved on the left hand side while the source/sink terms on the right hand side.


\bibliographystyle{ametsocV6}
\bibliography{references}

\end{document}